\documentclass[preprint,12pt]{elsarticle}

\usepackage{algorithmic}
\usepackage{multirow}
\usepackage{graphicx}
\usepackage{algorithm,algorithmic}
\usepackage{hyperref}
\hypersetup{hidelinks=true}
\usepackage{textcomp}
\usepackage{booktabs}
\usepackage{threeparttable}
\usepackage{fancyhdr}
\usepackage{amsmath,amssymb,amsfonts}
\usepackage{caption}   
\usepackage{subcaption} 
\usepackage{tabularx} 
\newtheorem{definition}{Definition}

\journal{Knowledge-based Systems}

\begin{document}

\begin{frontmatter}



\title{Explain EEG-based End-to-end Deep Learning Models in the Frequency Domain}


\author[1]{Hanqi Wang} 
\author[1]{Kun Yang}
\author[1]{Jingyu Zhang}
\author[2]{Tao Chen}
\author[1]{Liang Song\corref{cor}}
\cortext[cor]{Corresponding author}

\ead{songl@fudan.edu.cn}
\affiliation[1]{organization={Academy for Engineering and Technology, Fudan University},
            city={Shanghai},
            postcode={200433}, 
            country={China}}

\affiliation[2]{organization={School of Information Science and Technology, Fudan University},
            city={Shanghai},
            postcode={200433}, 
            country={China}}

\begin{abstract}
The recent rise of EEG-based end-to-end deep learning models presents a significant challenge in elucidating how these models process raw EEG signals and generate predictions in the frequency domain. This challenge limits the transparency and credibility of EEG-based end-to-end models, hindering their application in security-sensitive areas. To address this issue, we propose a mask perturbation method to explain the behavior of end-to-end models in the frequency domain. Considering the characteristics of EEG data, we introduce a target alignment loss to mitigate the out-of-distribution problem associated with perturbation operations. Additionally, we develop a perturbation generator to define perturbation generation in the frequency domain. Our explanation method is validated through experiments on multiple representative end-to-end deep learning models in the EEG decoding field, using an established EEG benchmark dataset. The results demonstrate the effectiveness and superiority of our method, and highlight its potential to advance research in EEG-based end-to-end models.
\end{abstract}




\end{frontmatter}



\section{Introduction}
\label{sec:introduction}
In the brain-computer interface research area, the electroencephalogram (EEG) serves as a reliable tool for measuring the electrical activity within the brain~\citep{r1}. Its utility spans various domains, such as emotion recognition~\citep{r15}. However, decoding EEG data is notably challenged by the low signal-to-noise ratio (SNR) \citep{r2,r3} and high dimensionality \citep{r1,r3}, impeding the application of traditional methodologies in this area. Consequently, there is a growing interest in leveraging deep learning techniques to handle complex EEG data due to its capacity to handle complex data. Recently, in numerous EEG-based applications, deep learning approaches have demonstrated superior performance compared to conventional methods~\citep{r4,r5,r6,r7,r8,r9}, thereby substantiating the efficacy of EEG-based deep learning models. Initially, deep learning models for EEG analysis usually relied on hand-craft features in frequency domain~\citep{r1,r3,r15}. This approach enhances the transparency and credibility of the models' decision-making processes, as previous studies have shown that the frequency domain contains indicative information relevant to various tasks~\citep{r10,r11,r12,r13}. But, the recent evolution towards end-to-end learning allows these models to learn directly from time-domain raw EEG data~\citep{r4,r5,r6,r7,r8,r9}. 

\begin{figure}[t]
\centerline{\includegraphics[width=\columnwidth]{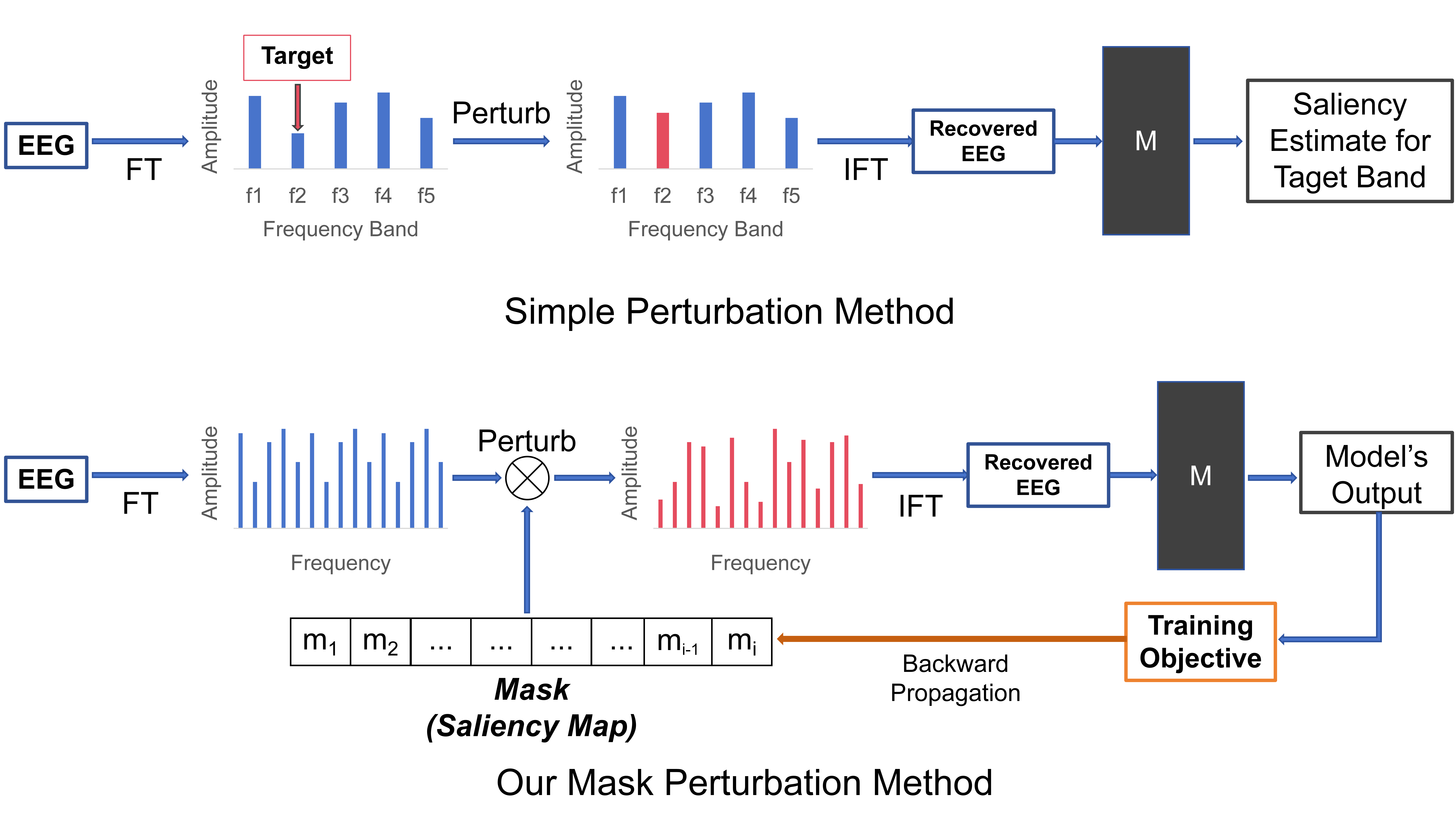}}
\caption{The illustration of the simple perturbation and ours. The first row shows the framework of simple perturbation, and the second row shows the framework of our method. For simplicity, this figure omits the perturbation generation parts and only emphasizes the mask mechanism. We can see that the simple perturbation can only estimate a specific band each time, without considering the joint role of each frequency component. The FT denotes the Fourier transform, the IFT denotes the inverse Fourier transform, and the M denotes the investigated model.}
\label{fig1}
\end{figure}

This shift shows the capability of deep learning to process raw EEG data, improving model accuracy and efficiency. Despite their success, the black-box nature of deep learning models poses a significant obstacle to understanding how the end-to-end models process the raw EEG signal and issue the predictions in the frequency domain. Since extensive research has revealed the relationship between EEG signals and brain activity in the frequency domain, improving decision-making process transparency in the frequency domain is critical to the credibility of these models. Thus, this problem hinders their deployment in security-sensitive domains. On the other hand, understanding the behaviors of these end-to-end models on the frequency domain also has the potential to facilitate the discovery of underlying cognitive processes~\citep{r14,r16,r23}. Thus, it is imperative to explain these end-to-end models that use raw EEG data as inputs. Such explorations are essential, yet current research in this domain remains limited, signaling a significant gap that motivates further investigation.

Explaining deep learning involves demonstrating how the deep learning model responds to the input features, with the saliency method being a prevalent technique. This approach highlights input regions critical to the model's prediction, providing insight into the model's decision-making process. Many existing works adopt it to explain the proposed end-to-end EEG-based model. For example, the authors in~\citep{r14,r17,r18,r19,r20} generate the saliency maps to illustrate the causal relationship of the investigated end-to-end models between the input and output. However, these works mainly focus on the time-domain explanation, failing to indicate the salient components in the frequency domain explicitly. To this end, some works~\citep{r21,r22,r23} have explored addressing this problem by perturbing at particular frequencies as shown in the top of Fig~\ref{fig1}. Instead of jointly generating a saliency map, these methods alter the features in a specific band at each operation to estimate the localized saliency, typically replacing the target band with random noise. 

Although this simple perturbation method demonstrated the effectiveness of investigating a specific band, they show limited ability to generate a global and accurate explanation for all the components in the frequency domain. The existing research in neuroscience has shown that there is a coupling between different frequencies of the EEG signal~\citep{r28,r29,r30}. Therefore, perturbing the EEG signal in a single frequency band alone is inadequate to model the dependence between frequencies. This limitation poses a challenge to the reliability of such an explanation method. Recent advances in other research fields proposed the mask perturbation technique to explain deep learning models~\citep{r14,r24,r25,r26,r27}. It adopts the mask to jointly perturb each component. By optimizing the mask, an accurate saliency estimation is given for each feature in the input. This technique has shown its ability to generate a global saliency map and model the inter-component dependency. The advantages of mask perturbation suggest its potential as a superior framework for explaining end-to-end EEG-based deep learning models in the frequency domain, which is shown at the bottom of Fig~\ref{fig1}. For this reason, this work aims to explore how to apply mask perturbation to explain end-to-end EEG-based deep learning model in frequency domain.

\begin{figure}[t]
\centerline{\includegraphics[width=\columnwidth]{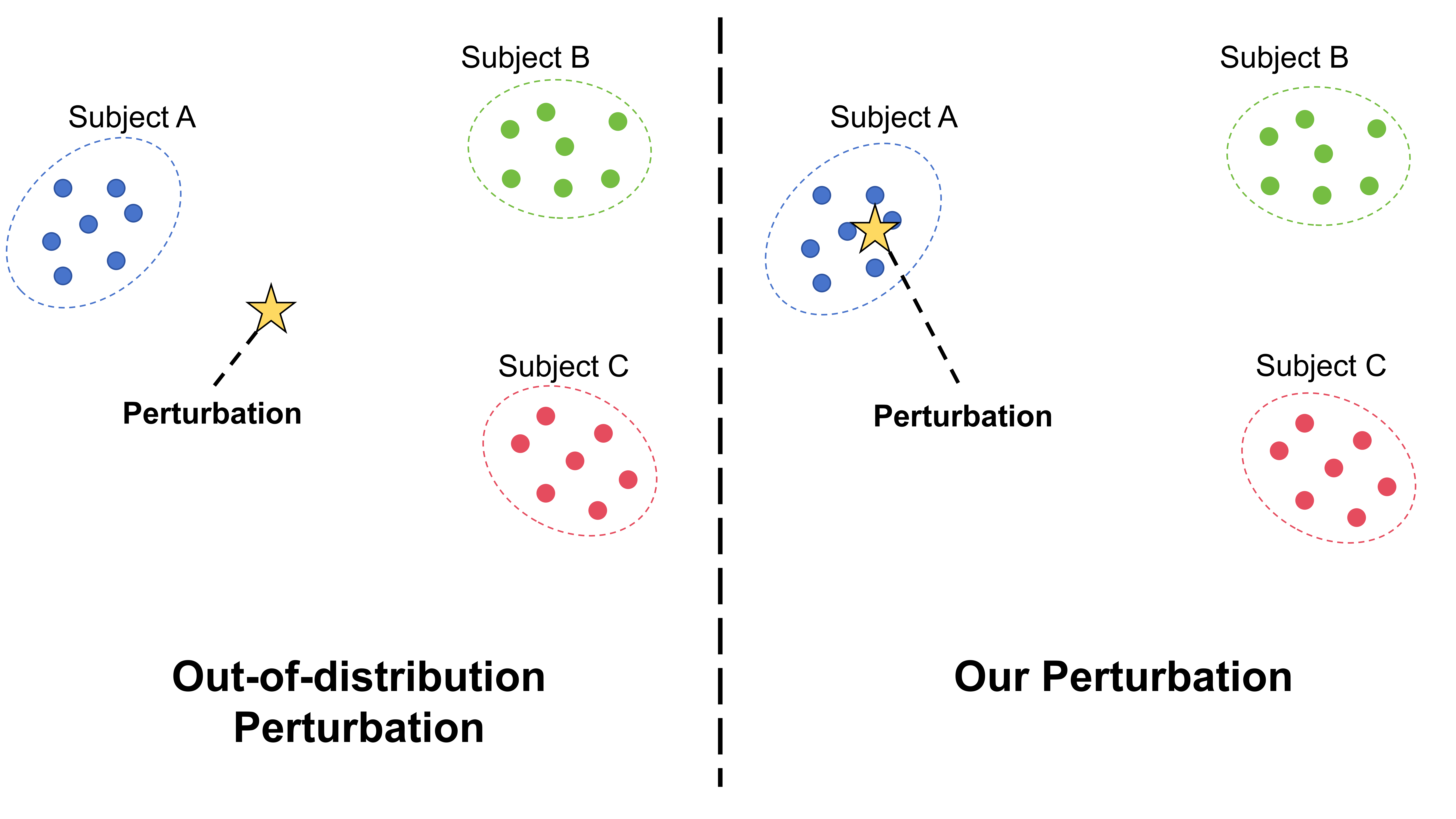}}
\caption{The illustration of distribution shift caused by perturbation. The left part demonstrates the out-of-distribution problem might caused by existing perturbation operation. The right part demonstrates data distribution of our perturbation method.}
\label{fig2}
\end{figure}

However, considering the characteristics of the EEG analysis research, our mask perturbation method faces two problems. 1) The perturbed EEG data is at the risk of triggering artifacts in the deep learning, potentially influencing the explanation~\citep{r24,r25,r32,r31,r33}. Specifically, the perturbation operation potentially causes the out-of-distribution sample. This deviation can lead to unexpected alterations in model output. Consequently, artifacts induced by out-of-distribution perturbations undermine the credibility and generalizability of the explanations. Therefore, it is crucial to encourage the perturbation to approximate the model's training data. Unlike other data types, EEG data exhibit significant inter-subject variance but maintain considerable intra-subject consistency~\citep{r1,r3,r15,r33}. As a result, EEG samples tend to cluster, showing significant discrepancies between different clusters. This observation suggests that the perturbation should be aligned with the data within a specific cluster to avoid the out-of-distribution problem. 2) The perturbation defined in the frequency domain is an open problem. Considering the coupling between frequency bands, the local or fixed perturbation is insufficient to determine the importance of a feature in the frequency domain. Instead, the perturbation in the frequency domain should capture the dynamic dependency between different frequency bands in a global view. Meanwhile, the frequency domain signal consists of real and imaginary parts, and the definition of the perturbation function needs to account for both real and imaginary information.

To address these issues, this work proposes a model-agnostic mask perturbation method to explain EEG-based end-to-end deep learning models in the frequency domain. Our approach leverages Fourier and inverse Fourier transformations to capture saliency for these time-domain EEG models in the frequency domain. To our knowledge, this is the first effort to introduce mask perturbation for explaining EEG-based models in the frequency domain. Considering the risk of triggering artifacts, our method proposes a target alignment loss to avoid the out-of-distribution problem. The perturbations are encouraged to approach a target cluster in training data, aligning the perturbation distribution. To generate perturbation for EEG data in the frequency domain, we propose a perturbation generator network to learn the perturbation. This network determines the saliency of a feature in the frequency domain using global information. And, it consists of two branches, real and imaginary, to learn from complex spectrum data. This design enhances the perturbation's capability to capture dynamic dependency and utilize both real and imaginary parts in the frequency domain. In the experiment, our method is evaluated on multiple representative EEG-based deep learning models and a well-established EEG benchmark dataset. The performance of our method supports the significance of our method in identifying the salient frequencies. And, we compare the previously proposed simple perturbation with our method. The outcome shows that our method performs better than its efficiency and accuracy. In addition, to show the potential value of our research, we implement an experiment to investigate the difference between subject-independent models and the subject-dependent models in the frequency domain. Finally, we implement an ablation study on the indispensability of each proposed module. The contribution of our work can be summarized as follows:
\begin{enumerate}
\item We propose a mask perturbation framework to explain the end-to-end EEG-based deep learning model in the frequency domain. Our method adopts Fourier and inverse Fourier transmissions to address the problem of capturing the saliency for the time-domain model in the frequency domain.

\item We propose a target alignment loss to address the out-of-distribution problem, enhancing the reliability of the generated explanation. Our method encourages the perturbation to approach the target cluster samples, aligning the perturbation distribution.

\item We propose a frequency-domain perturbation generator to define the perturbation. It adopts the global information to capture dynamic dependency and two-branch structure to include information from both real and imaginary parts.

\item We implement comprehensive experiments to validate our method and exemplify the application of our research. The results show the effectiveness of the whole method and each proposed strategy. We further implement an example experiment to analyze the relationship between frequency and generalization behavior in an EEG-based deep learning model.
\end{enumerate}

\section{Related Works}
\subsection{Saliency Method for Explaining EEG-based Deep Learning Models}
Gradient-based saliency map method is extensively employed to improve the transparency for EEG-based deep learning model. This method traces gradients from the model’s output back to its input. The underlying assumption of gradient-based methods is that a salient feature is able to drastically impact the model's prediction when it is changed. Accordingly, gradients signify sensitivity to these changes, aligning with the definition of saliency. Many previous studies \citep{r17,r18,r19,r20} have utilized the gradient-based approach outlined in \citep{r21} to generate saliency maps for EEG-based deep learning models. Moreover, some recent advances based on this technique are proposed to explain deep learning model. For example, gradient-class activation mapping (Grad-CAM) in \citep{r35} utilize class-specific gradients at the last convolutional layer to identify salient areas of the input. Some recent works for decoding raw EEG data introduce this approach to improve their transparency~\citep{r36,r37,r38}. Despite these innovations, all gradient-based approaches share a common limitation. These works capture the averaged property of the model, showing limited ability in understanding the information in the input instances~\citep{r24,r14}. More importantly, there is still an absence of extension for this approach to explain the end-to-end EEG-based model in the frequency domain. On the other hand, the perturbation-based method has gained popularity for explaining deep learning models. This approach evaluates how changes to the input affect the output to determine saliency~\citep{r39}. A region is regarded as salient if its perturbation significantly impacts the output. Recently, this method has been introduced to explain the EEG-based deep learning model. For example, the authors in~\citep{r14} proposed a context-aware perturbation method to generate the saliency map for EEG data. While the inclusion of contextual information in perturbation alleviates the risk of artifacts, it cannot directly reveal the behaviors of end-to-end EEG-based deep learning models in the frequency domain. To address this problem, some works~\citep{r21,r22,r23} have explored addressing this problem by perturbing at particular frequencies. For example, easyPEASI proposes replacing the real and imaginary parts in the target band with Gaussian noise~\citep{r23}. However, this approach is insufficient for modeling the joint effect of various frequency components. Additionally, noise-based perturbation cannot guarantee that the perturbation is uninformative or alleviate the out-of-distribution problem. Moreover, this simple perturbation method is limited by extensive computation when extended to generate instance-level saliency maps~\citep{r14}.

\subsection{Mask Perturbation Method}
The mask perturbation-based method, initially detailed in references \citep{r24,r25,r26}, offers a refined approach to generating saliency maps for image data. This method is built based on the perturbation-based method, which assumes the features sensitive to alternation are salient. This method has shown its advantages in several key research areas. Firstly, it provides a more targeted detection of saliency at the level of individual inputs, as analyzed by the authors in \citep{r24, r26}. Unlike gradient-based methods, mask perturbation directly interacts with the input through a mask of identical dimensions, facilitating an element-wise analysis that assesses the entire input comprehensively. Secondly, this approach allows for the exploration of global impacts, offering a broader view of how perturbations affect model behavior. Thirdly, it defines the generation of explanations as a learning task, employing a meta-predictor to derive precise and reasonable estimates for saliency. This method has proven highly effective in exposing the workings of deep learning in many research fields, including time series analysis~\citep{r27,r40,r41,r14,r46,r48}. For example, the authors in~\citep{r46} proposed learning perturbation to explain time series prediction. However, their learning learning method, focused on the time domain, shows limited consideration for processing complex spectrum data. And, Dynamask in~\citep{r14} utilizes the local adjacent information to determine the perturbation. Although recent works have advanced the field of explaining time series models, there remains a gap in extending these explanations to EEG-based end-to-end deep learning models in the frequency domain. In this work, we aim to extend the mask perturbation method to address this gap, providing a robust framework for frequency-domain explanation.

\section{Methodology}
\begin{figure}[ht]
\centerline{\includegraphics[width=\textwidth]{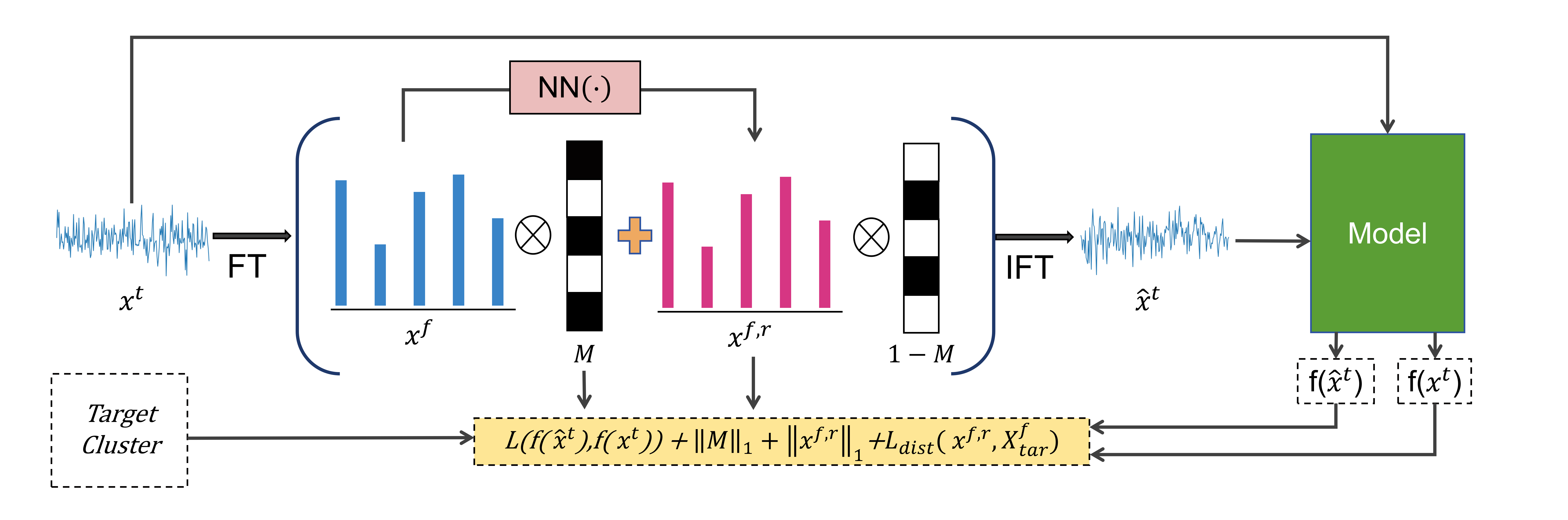}}
\caption{ The overall structure of the proposed explaining method.}
\label{fig3}
\end{figure}

In this section, the proposed explaining method is illustrated in detail. The overall structure can be seen in the Fig.~\ref{fig3}. First we introduce the basic concept of mask perturbation in our scenario, including domain conversion, mask mechanism in frequency domain, and domain inverse conversion. And then, we illustrate the perturbation operation. Following it, the proposed perturbation generator network is presented. Next, the proposed target alignment loss is introduced to demonstrate how the perturbation is encouraged to approach the normal data. Finally, we discuss the optimization strategy of the proposed method. 

\subsection{Mask Mechanism}
This subsection describes how we perform mask perturbation on frequency domain. In this paper, we use $x^t$ to denote the investigated raw EEG sample, ${x^t}\in R^{Ch\times T}$. The $T$ denotes the time steps, and $Ch$ denotes the number of channel. For raw EEG data, each channel contains the EEG signal collected from a electrode, i.e., the brain activities in a certain brain area. To convert the raw EEG data to frequency domain, we apply fast Fourier transform to the raw EEG data, obtaining the frequency-domain sample ${x^f}$. Before the discussion of mask perturbation, we first formalized the definition of the mask in this work as the follows:

\begin{definition}
A mask $M$ is a learnable matrix bound with the given sample $x^f$ and model $f$. The $M$ in our method serves as both the perturbation operator and saliency map result. Each element in the $M$ is within the range $[0,1]$, indicating the saliency of the corresponding region in the sample. For a given element $m$ in $M$, its corresponding region in the sample is regarded as salient if $m$ is close to 1. Otherwise, the corresponding region in the sample would be regarded as the opposite. 
\end{definition}

This structure allows $M$ to directly represent the saliency map for a specific sample $x^f$. It offers a clear visualization of which parts of the $x^f$ are considered salient by the model $f$. This alignment between $M$ and $x^f$ facilitates an intuitive and precise method for identifying the salient features in the sample. In practice, researchers may be more interested in the saliency of frequency bands rather than individual frequencies. For this reason, we adopt a scalable mask mechanism to perturb the sample. The length of learnable mask is the factor of the length of sample $x^f$. When performing perturbation operations, we use repeat extensions to ensure that the inputs and masks are of the same size.

In our method, the search for the optimal mask is formulated as a machine-learning task. And, the mask $M$ dictates how to perturb the sample as follows:
\begin{equation}
\hat{x}^f=x^{f} M + (1-M)x^{f,r}.
\label{eq1}
\end{equation}
Let $\hat{x}^f$ represent the perturbed sample in the frequency domain, and $x^{f,r}$ denote the perturbation on the sample $x^f$, which replace the original sample under the control of $M$.

\subsection{Perturbation Generator Network}
\begin{figure}[t]
\centerline{\includegraphics[width=\columnwidth]{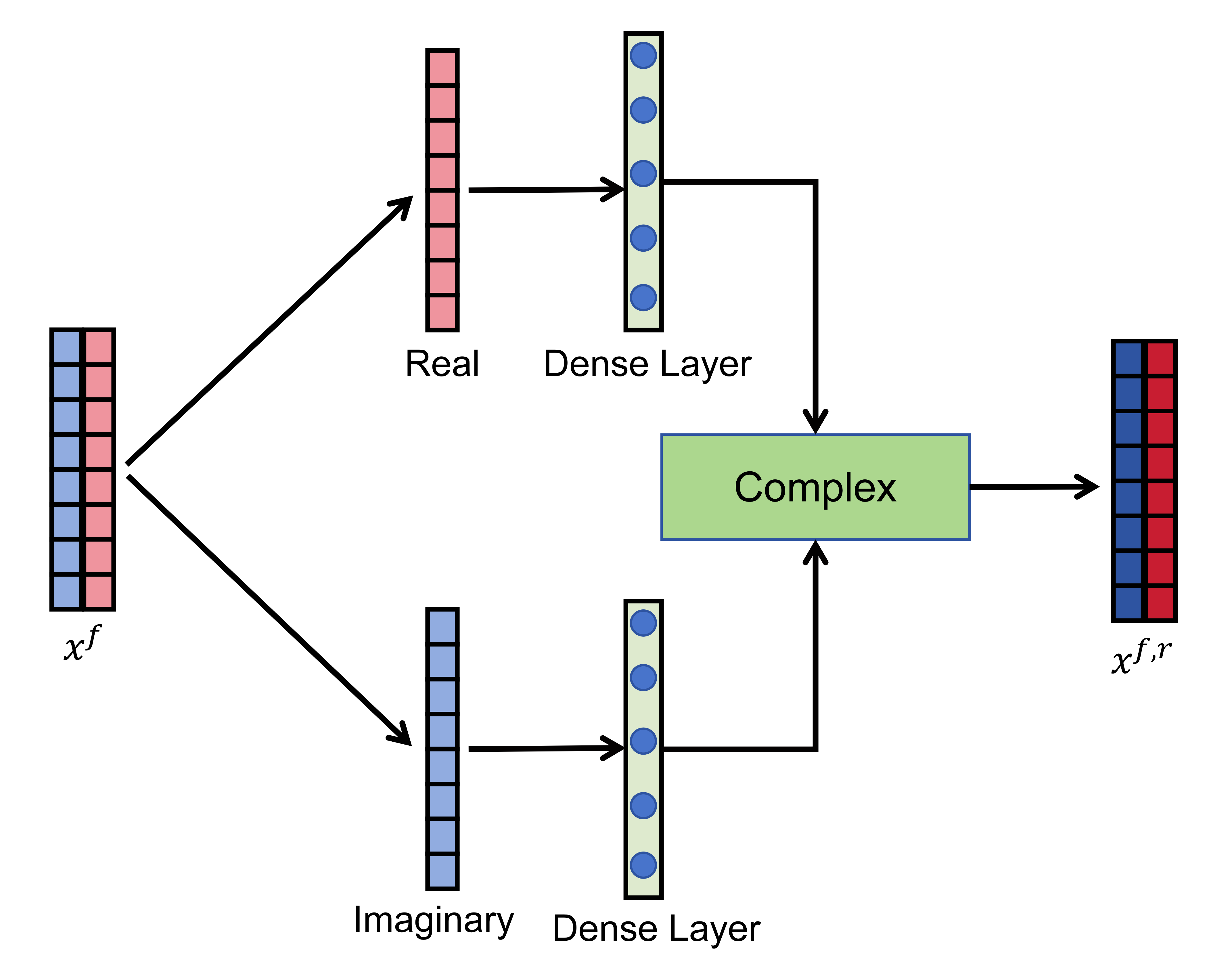}}
\caption{The structure of perturbation generator network $NN(\cdot)$. The $Complex$ denotes the complex features construction.}
\label{fig4}
\end{figure}
Many existing works rely on fixed and local information to generate the perturbation, capturing temporal dependency and avoiding the out-of-distribution problem~\citep{r46,r14,r27}. For example, the work in~\citep{r14} utilizes the local context information to determine the perturbation. However, these approaches show limited ability to dynamically capture the global dependency, misleading the saliency search. For this reason, we introduce a learning method to generate the perturbation, as shown in the Fig.~\ref{fig4}.

This network uses the dense layer to learn the dynamic dependency between the different frequency components in the sample. Thus, it can determine the perturbation of a certain frequency component from a global perspective in the frequency domain. And, this neural network adopts a two branches architecture. The perturbation learning can be formulated as the~\eqref{eq2}.
\begin{equation}
\begin{aligned}
    &x^{f,r}=NN(x^{f});\\
    &x^{f,r}_{real}=Dense(Real(x^{f})), \\
    &x^{f,r}_{img}=Dense(Img(x^{f})),\\
    &x^{f,r}=Complex(x^{f,r}_{real},x^{f,r}_{img}),
    \label{eq2}
\end{aligned}
\end{equation}
 where $NN(\cdot)$ denotes the perturbation generator network, $Complex$ denotes the construction of a complex spectrum with real and imaginary parts, and $Dense(\cdot)$ denotes the dense layer. The real branch learns the perturbation in the real part, and the imaginary branch learns the perturbation in the imaginary part. Finally, we recombined the real and imaginary parts to form a complex frequency domain perturbation.

\subsection{Target Alignment Loss}
EEG samples are prone to exhibit high inter-subject variability and intra-subject consistency. This causes data samples within the same subject to cluster closely together, with significant discrepancies between different clusters. For the reliability of the generated saliency map, the perturbations must approximate the cluster of a specific training subject. Otherwise, the perturbation might present a risk of triggering artifacts due to the out-of-distribution problem. Therefore, in the first step, we need to determine which cluster the perturbation should approach. Suppose $j_{th}$ subject cluster has $N_j$ data points, we calculate the center for each training subject as follows:
\begin{equation}
    C_j = \frac{1}{N_j} \sum_{i=1}^{N_j} x^t_{j,i},
    \label{eq3}
\end{equation}
$x^t_{j,i}$ denotes the $i_{th}$ samples collected from the subject $j$, and $C_j$ denotes the center point of the subject $j$. And then, we select the cluster whose center is closest to the ${x}^{t}$ as the target cluster. To reduce the shift caused by perturbation, a target alignment loss is proposed to explicitly bring the perturbation closer to all samples within the target cluster in the frequency domain. This constraint imposed on the mask search process is given by: 
\begin{equation}
    L_{dist} = \frac{1}{N_{tar}} \sum_{i=1}^{N_{tar}} ({x}^{f,r}-x^f_{tar,i}),
    \label{eq4}
\end{equation}
where $x^f_{tar,i}$ denotes the $i_{th}$ samples in the target cluster in the frequency domain, ${x}^{f,r}$ denotes the perturbation in the frequency domain obtained, and $N_{tar}$ denotes the samples of the target cluster.

\subsection{Mask Optimization}
Following perturbation, we apply an inverse fast Fourier transform to $\hat{x}^f$, resulting in the recovered input in the time domain $\hat{x}^t$. For a black-box model $f(\cdot)$, its prediction for the perturbed sample, $f(\hat{x}^t)$, is generated by feeding $\hat{x}^t$ into the model. And, the prediction for the original sample, $f({x}^t)$, is generated by feeding ${x}^t$ into the model. Our method operates under the assumption of the perturbation-based method that the model's output is sensitive to perturbations in salient regions while remaining robust against perturbations in non-salient regions. Based on this assumption, the optimization of our mask follows the preservation game strategy in~\citep{r27,r46}, aiming to mask as much data as possible while preserving the as close predictions as possible to the original ones. Our method's fundamental optimization objective aims to minimize discrepancies between $f(\hat{x}^t)$ and $f({x}^t)$, i.e., minimizing $L(f(\hat{x}^t), f({x}^t))$. The $L(\cdot)$ depends on the task of the model, e.g., cross-entropy loss for the classification task. Under this strategy, our optimization objective guides the mask to preserve salient regions by setting corresponding mask elements close to 1 and suppresses non-salient regions by setting corresponding mask elements close to 0.

However, the optimization of mask may take shortcuts, such as making $m=1$, or $m=0$ and $NN(x^f)=x^f$. In this way, the predictions can be the same with the original ones, while the mask fails to capture the saliency distribution. Thus, we need to prevent this behavior. Following the strategy in~\citep{r46}, the $\Vert M \Vert_1$ and $\Vert {x}^{f,r} \Vert_1$ are added in the training objective as regularized term. Finally, the training objective of our mask search is given by:
\begin{equation}
     L(f(\hat{x}^f), f({x}^f))+ \Vert M \Vert_1 + \Vert {x}^{f,r} \Vert_1 + \lambda L_{dist}.
    \label{eq5}
\end{equation}
The $\Vert M \Vert_1$ inhibits the preservation of original sample and encourage the perturbation to replace it. Moreover, the $\Vert {x}^{f,r} \Vert_1$ entices the ${x}^{f,r}=NN(x^f)$ to be uninformative. The $\lambda$ is a hyper-parameter.

\section{Experimental Results and Analysis}
\subsection{Dataset and Model Introduction}
we conduct experiment on a public EEG dataset SEED \citep{r10}, which is commonly used in emotion recognition task. The SEED~\citep{r36} comprises EEG signals collected from 15 participants, categorized into three emotional states: positive, neutral, and negative. Each participant completed three sessions, each consisting of 15 trials conducted at different times. The EEG signal is recorded at sampling rate 200 Hz. In the experiment, we perform a Pos-Neg binary classification, using non-overlapping 2-s segments.

\begin{table}[!ht]
    \caption{The number of trainable parameters of the three selected models.}
    \centering
    \begin{tabular}{|c|c|}
    \hline
    Model              & Parameters   \\
    \hline
    EEGNet            & 2578\\
    \hline
    Tsception         & 58178\\
    \hline
    DeepConvNet       & 171702\\
    \hline
    \end{tabular}
    
    \label{tab1}
\end{table}
To validate the performance of our method, we carefully selected three models: one state-of-the-art deep model for emotion recognition, Tsception~\citep{r20}, and two commonly used general models, EEGNet~\citep{r8} and DeepConvNet~\citep{r21}. The selection of these models is based on two considerations. First, we believe our method should be tested on network designs with various target purposes. Second, we are interested in assessing the effectiveness of our method on models of different scales. As shown in Table \ref{tab1}, these three models demonstrate an increasing order of magnitude in the number of trainable parameters. All three models operate end-to-end, taking raw EEG data as input and outputting scores for each class. They are implemented on the SEED dataset to perform binary classification task, following the same paradigm.

\subsection{Experiment Settings}
The three models are all trained with the default settings reported in the original papers. All the experiment are implemented on a GTX 2080Ti GPU. For the training of mask, we set the hyperparameters $\lambda$ as 0.05. Then, we use the cross-entropy loss to measure the discrepancy between perturbed predictions and original ones. And, the length of scalable mask is set to 10 on frequency dimension. We adopt Adam as an optimizer, with a learning rate equal to 0.01. Moreover, each mask is trained with 300 epochs. To improve the efficiency, an early stopping strategy, which would break training after 10 epochs if there is no significant reduction in loss, is adopted in the training process.
\begin{figure*}[htbp]
    \centering
    \begin{subfigure}[b]{0.32\linewidth}
        \centering
        \includegraphics[width=\linewidth]{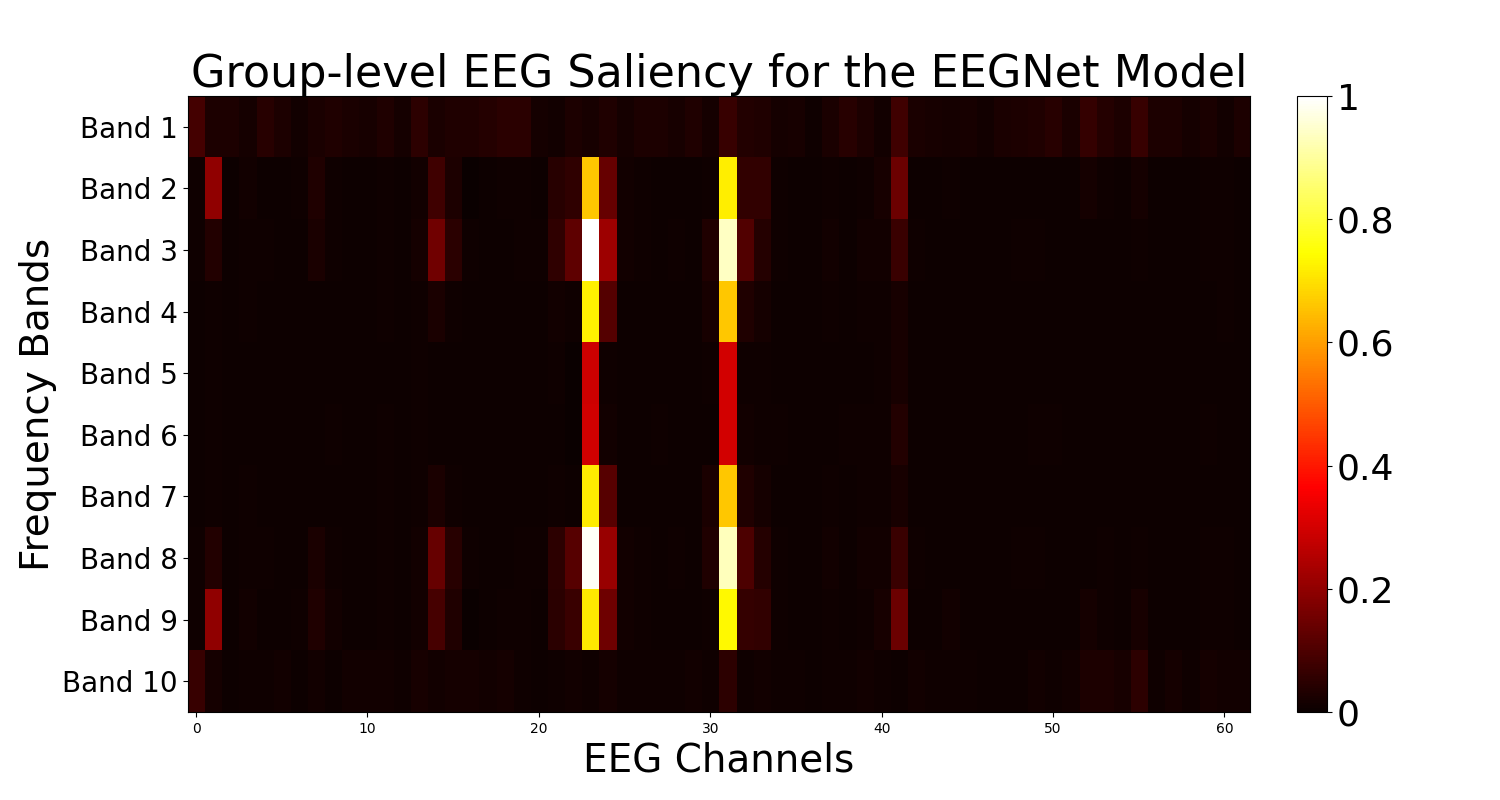}
        \label{fig5_1}
    \end{subfigure}
    \hfill
    \begin{subfigure}[b]{0.32\linewidth}
        \centering
        \includegraphics[width=\linewidth]{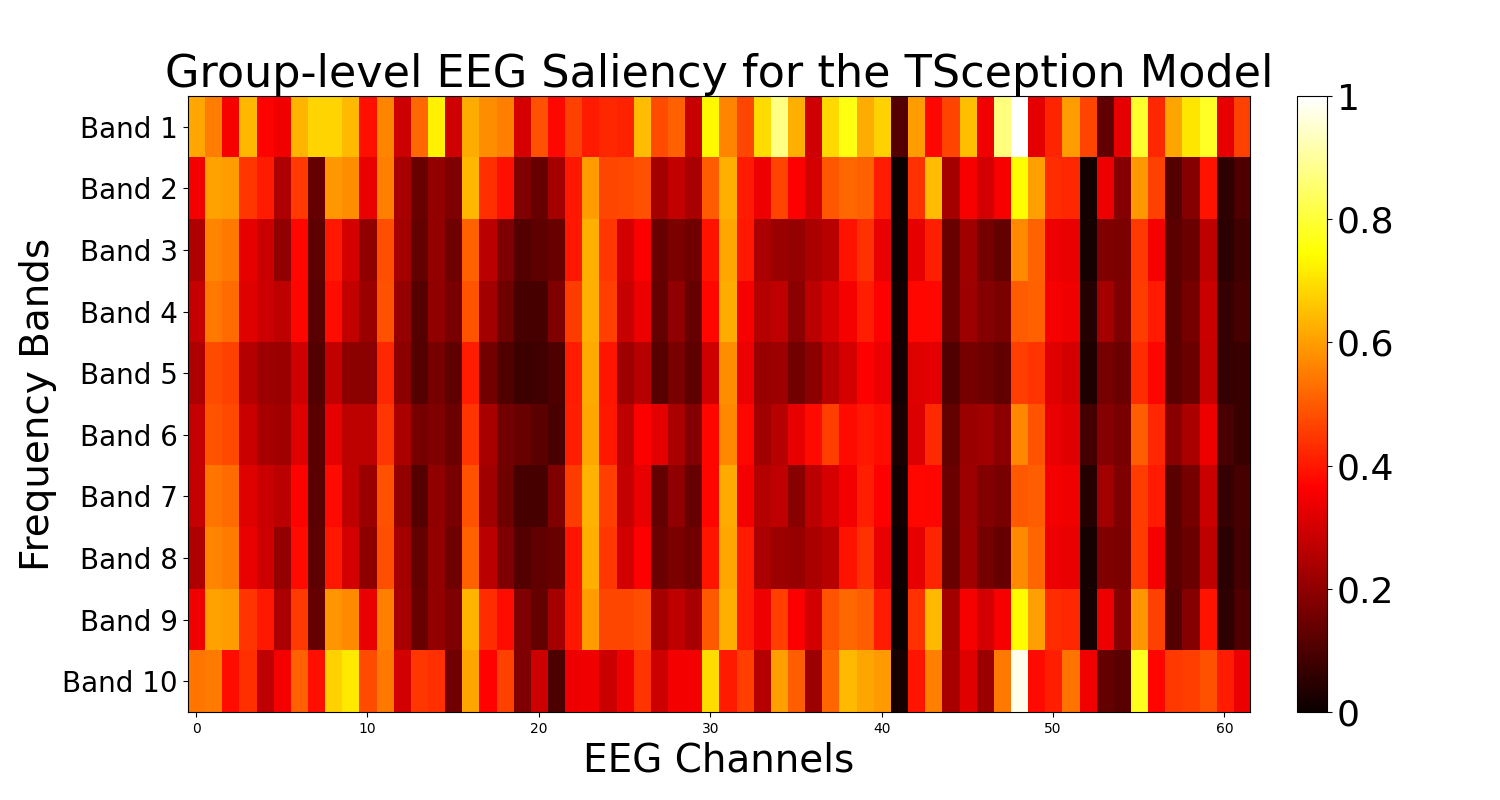}
        \label{fig5_2}
    \end{subfigure}
    \hfill
    \begin{subfigure}[b]{0.32\linewidth}
        \centering
        \includegraphics[width=\linewidth]{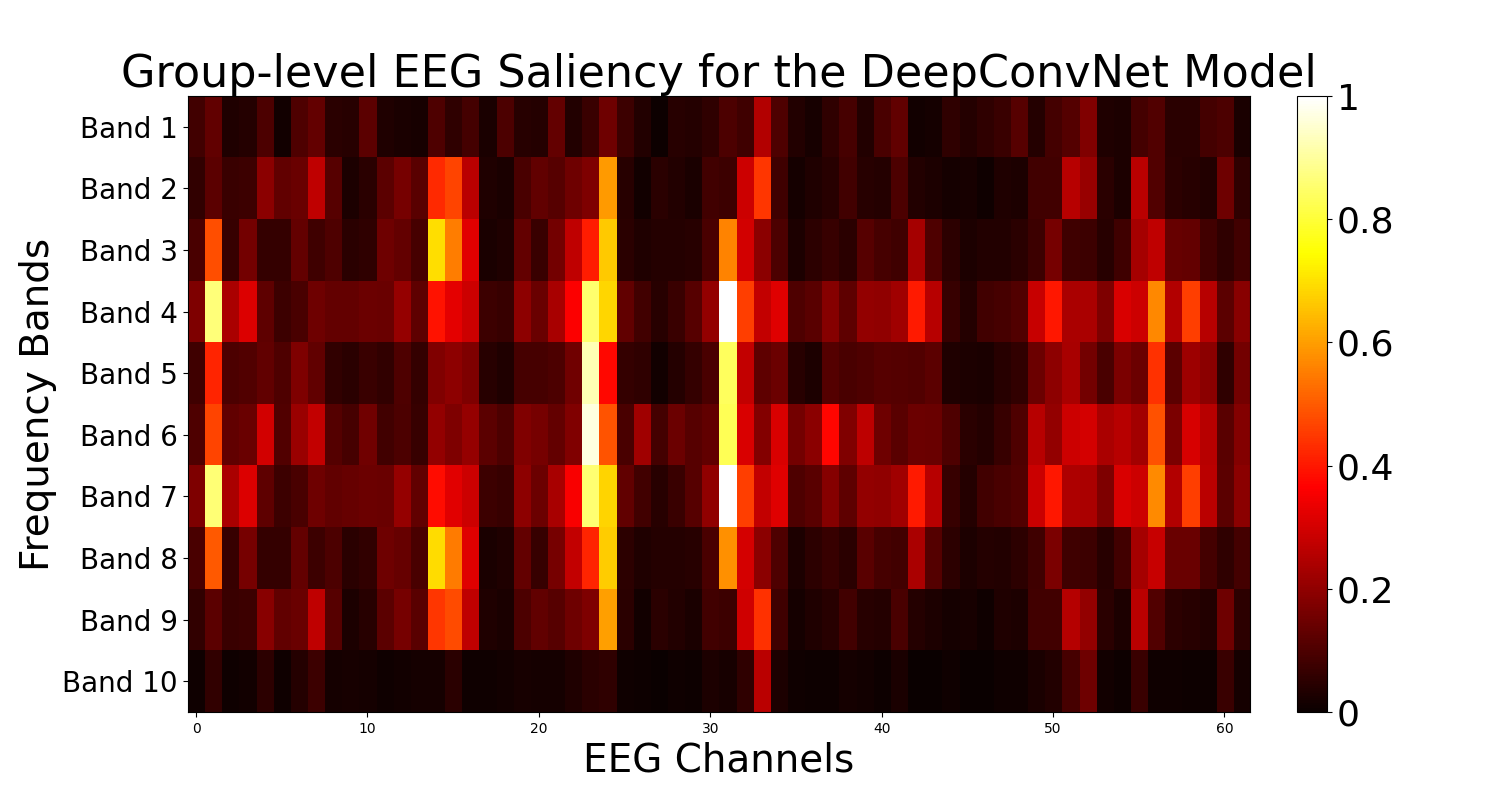}
        \label{fig5_3}
    \end{subfigure}
    \caption{The averaged saliency map in frequency domain for the three selected model using our method.}
    \label{fig5}
\end{figure*}

\begin{figure*}[htbp]
    \centering
    \begin{subfigure}[b]{0.32\linewidth}
        \centering
        \includegraphics[width=\linewidth]{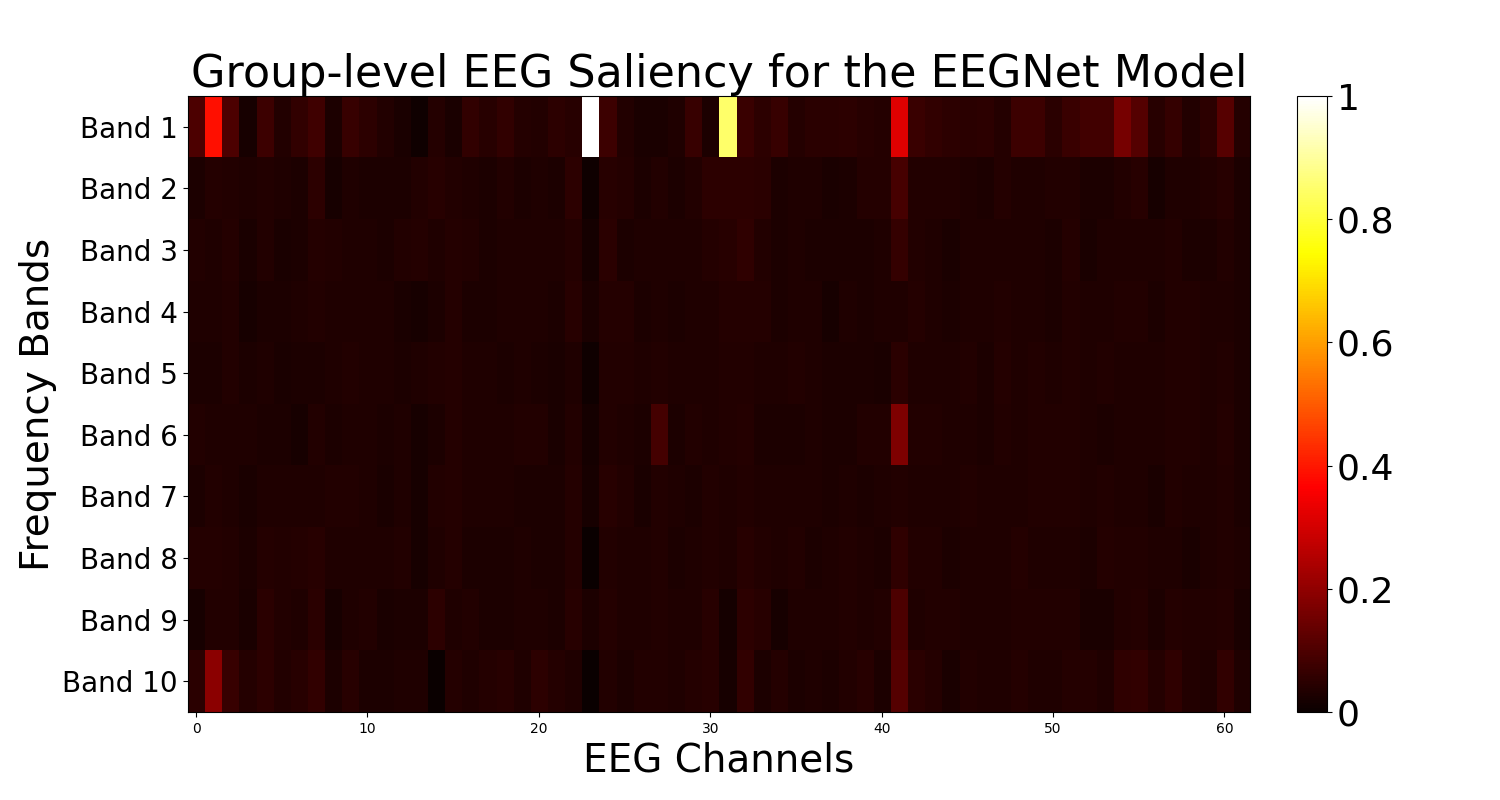}
        \label{fig6_1}
    \end{subfigure}
    \hfill
    \begin{subfigure}[b]{0.32\linewidth}
        \centering
        \includegraphics[width=\linewidth]{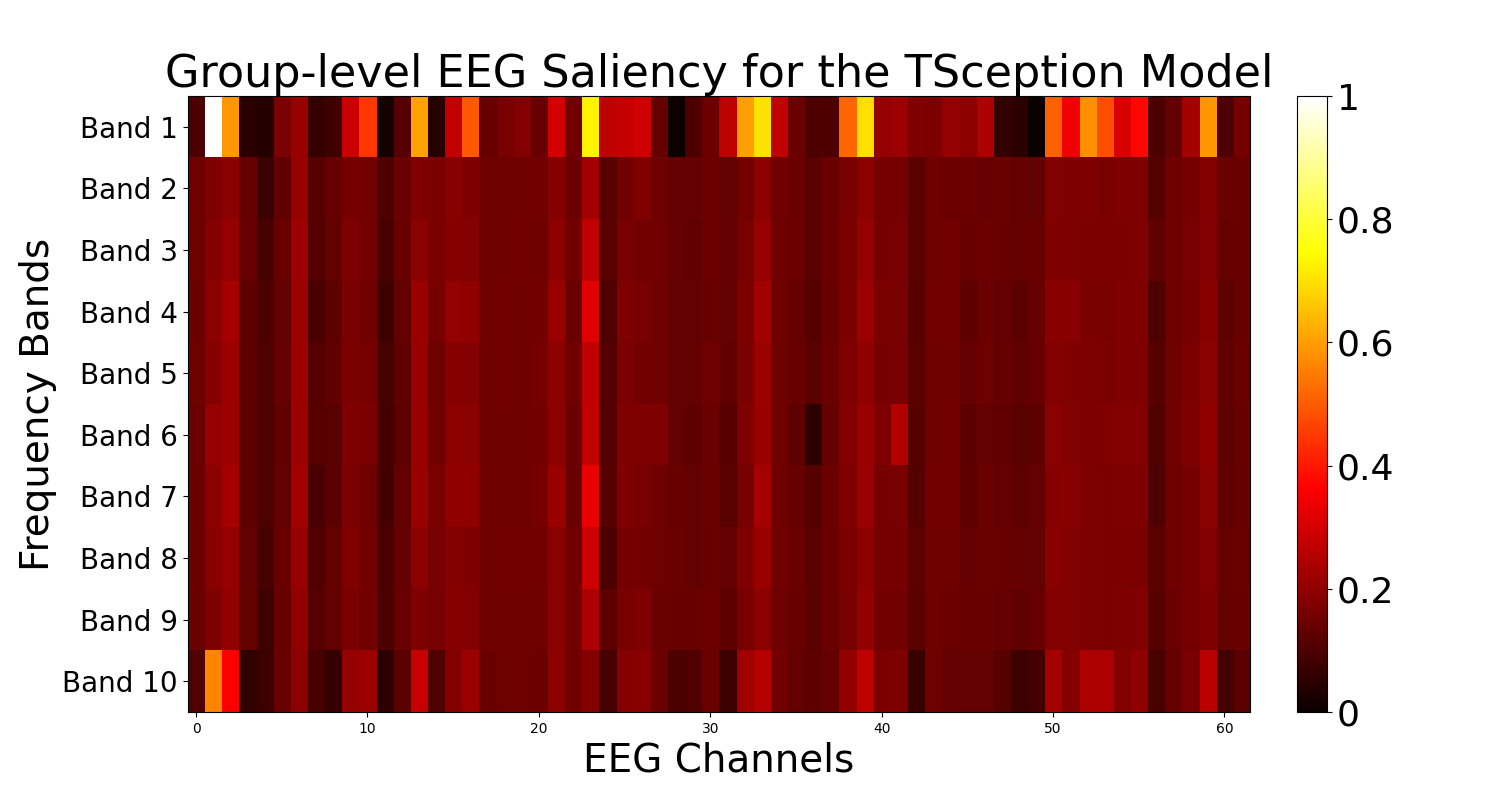}
        \label{fig6_2}
    \end{subfigure}
    \hfill
    \begin{subfigure}[b]{0.32\linewidth}
        \centering
        \includegraphics[width=\linewidth]{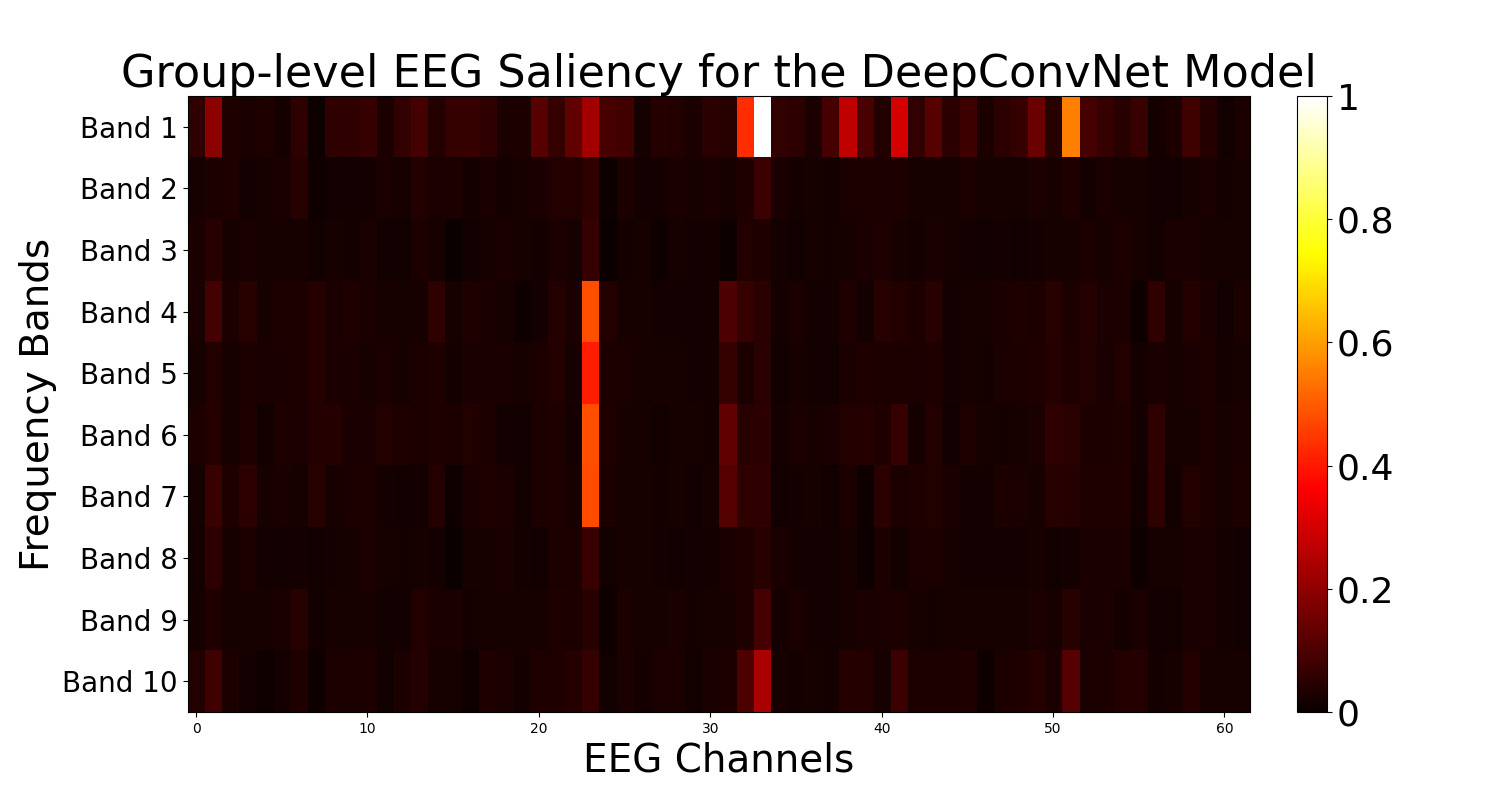}
        \label{fig6_3}
    \end{subfigure}
    \caption{The averaged saliency map in frequency domain for the three selected model using simple perturabtion.}
    \label{fig6}
\end{figure*}
\subsection{Group-level Saliency Validation}
On the SEED dataset, we adopt the well-established subject-independent protocol to train the three models, using the hyperparameters reported in the original papers. The subject-independent training protocol in EEG research involves training models on data from multiple individuals to develop generalized patterns, thereby enhancing the model's generalizability and robustness across diverse subjects. In our experiment, we use data from all subjects to train the models.

Understanding the overall model’s behavior across groups is of interest in many areas~\citep{r42,r43,r44,r45,r48}. Considering this, the proposed method is validated for its effectiveness in capturing group-level saliency. To obtain the statistic of the group-level saliency, we average the masks over all samples—the obtained mask comprises channel and frequency dimensions. The obtained saliency map for each band in each channel is presented in Fig.~\ref{fig5}. Following the works in~\citep{r14,r17}, we validate the effectiveness of captured saliency using a removal and feed-in game.

First, we divide the elements in the mask into salient and non-salient categories, using the median of the elements in the average mask as the threshold. Second, we remove the frequency components corresponding to the salient and non-salient elements separately by setting them to zero. Following it, the perturbed sample is then transformed back into the time domain using an inverse Fourier transform, and the recovered EEG signal is fed into the end-to-end model. Finally, the observed drop in performance is used to measure the saliency of the removed components.

To compare our work with the existing work, we re-implement the state-of-the-art frequency-domain simple perturbation method, easyPEASI, proposed in~\citep{r23}. We iteratively replace the real and imaginary parts of each band in each channel with Gaussian noise. Then, we use the performance drop to generate the saliency map, as shown in Fig.~\ref{fig6}. Similar to what we describe above, the saliency map generated by eastyPEASI is also used as evidence to measure the removed components.  We can observe that the saliency of the spectrum in Fig.~\ref{fig5} is symmetric along the frequency band axis, due to the symmetry of the spectrum on the frequency axis. In contrast, the saliency map of simple perturbation does not exhibit strict symmetry. This observation implicitly confirms the effectiveness and reasonableness of our saliency approach. 

\begin{figure}[htbp]
    \centering
    \begin{subfigure}[b]{\linewidth}
        \centering
        \includegraphics[width=\linewidth]{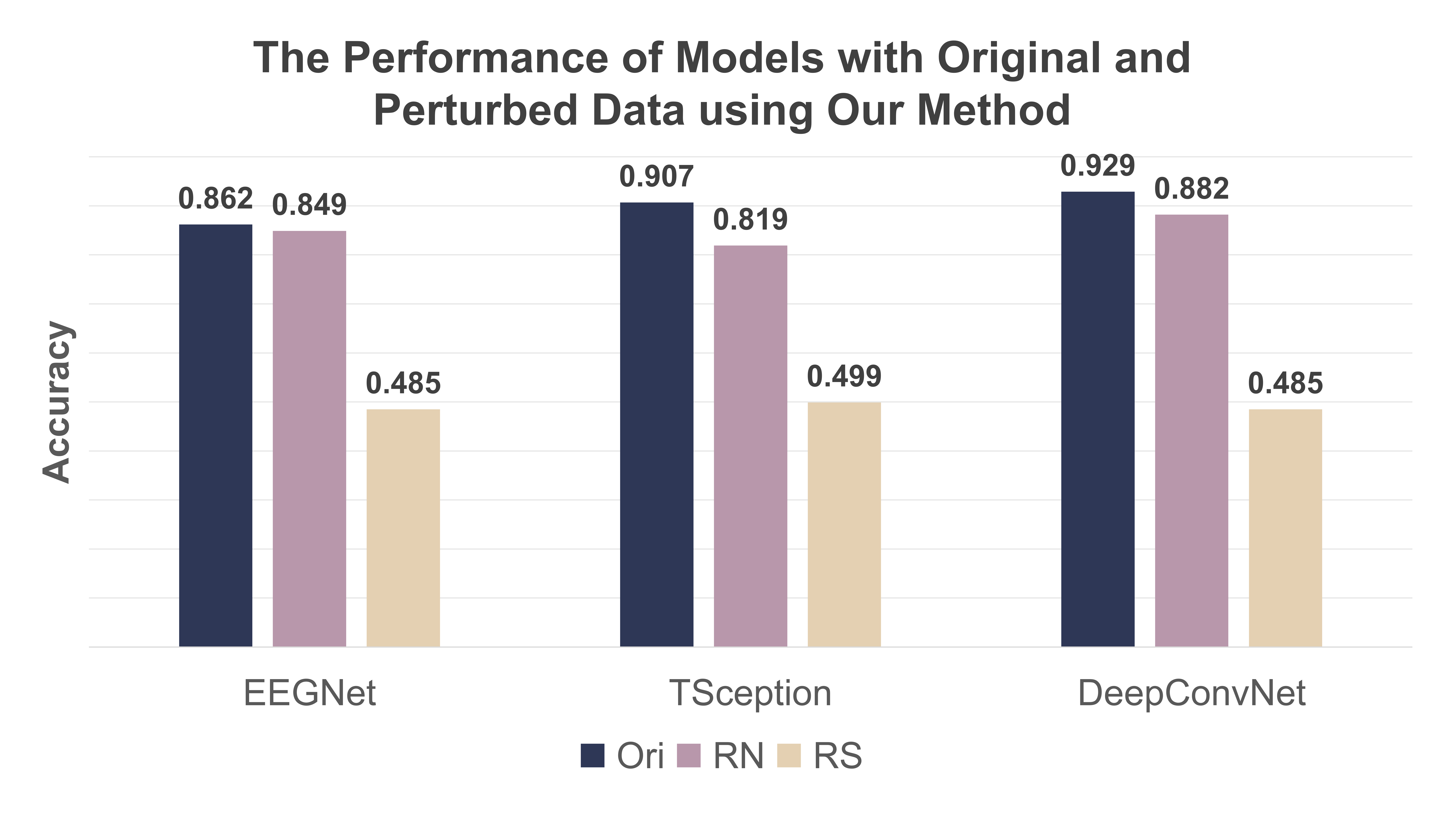}
        \label{fig5_1}
    \end{subfigure}
    \hfill
    \begin{subfigure}[b]{\linewidth}
        \centering
        \includegraphics[width=\linewidth]{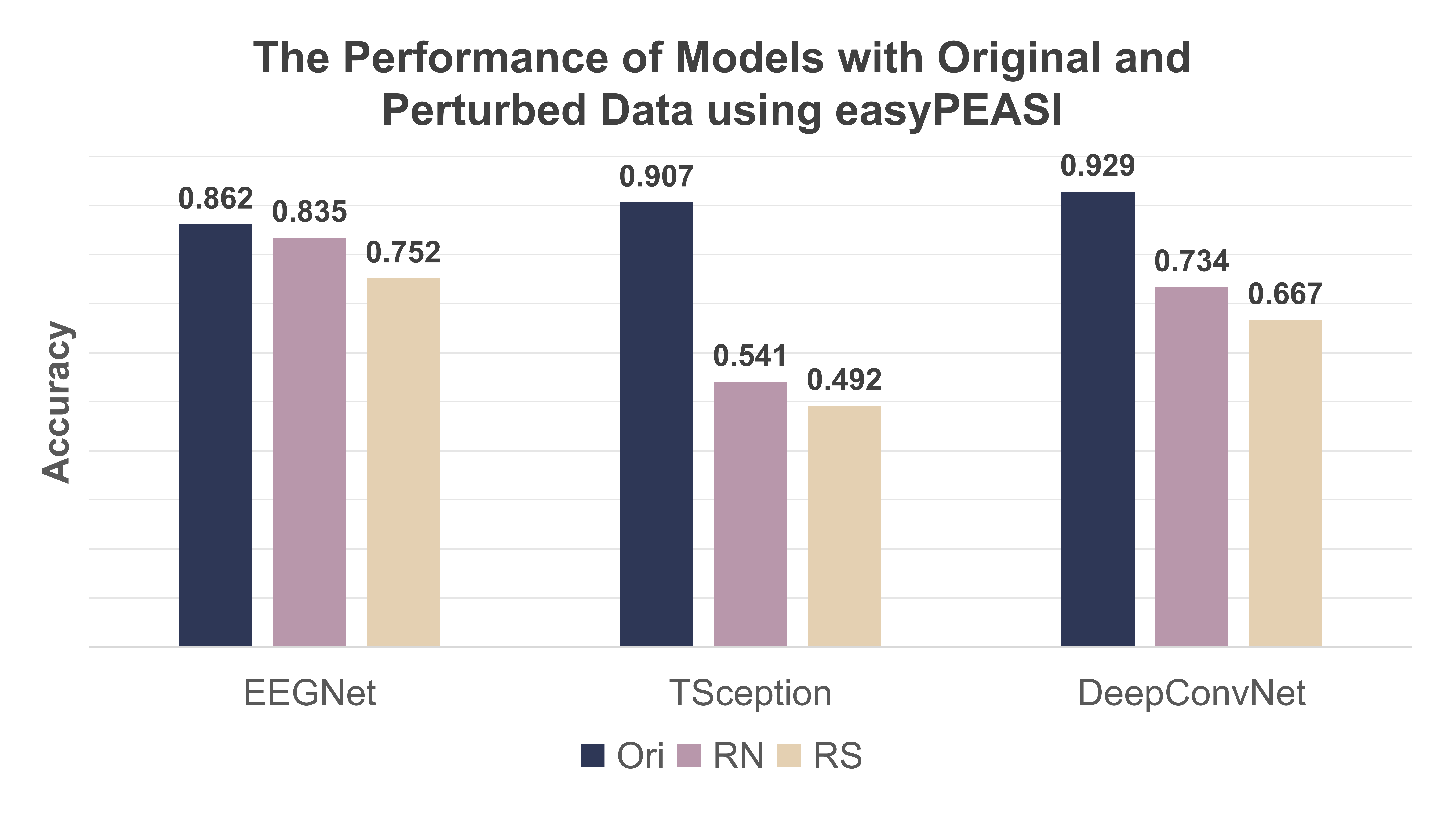}
        \label{fig5_2}
    \end{subfigure}
    \caption{The comparison of our method and simple perturbation, easyPEASI, is shown in the figure. The accuracies of models using different input data are listed in the fire. Ori: original data; RN: removing non-salient components; RS: removing salient components.}
    \label{fig7}
\end{figure}

According to the definition, saliency represents the impact of corresponding components in the data on the model's prediction. Thus, capturing saliency can be considered effective if removing components of the data that correspond to salient elements in the mask has a significantly greater impact on performance, while removing components that correspond to non-salient elements has a significantly weaker impact on performance. The comparison of our method and easyPEASI is shown in Fig.~\ref{fig7}. As observed, the performance drops using our method are substantial across all three models, whereas the drops using easyPEASI are non-significant. This demonstrates that easyPEASI has limited capacity in capturing saliency compared to our method.

\subsection{Instance-level Saliency Validation}
Although the demand for instance-level saliency is not as widespread as group-level saliency, capturing instance-level saliency can further demonstrate the effectiveness of interpretation methods and meet potential research needs. Therefore, here we further validate the capability of the proposed method to capture instance-level saliency. Additionally, one advantage of our method is its ability to capture instance-level saliency at a relatively low cost compared to simple perturbation.

Existing research has demonstrated that simple perturbation is constrained by extensive computational requirements~\citep{r14}. Suppose we have \(N\) samples with a shape of \(C \times F\); simple perturbation would require \(N \times C \times F\) perturbation operations and model inferences to determine the saliency for each component in every sample. This is impractical, especially when the data size is large. Thus, our mask perturbation method not only improves the effectiveness of saliency capture but also facilitates the extension of frequency domain explanations to the instance level.
\begin{figure}[ht]
\centerline{\includegraphics[width=\columnwidth]{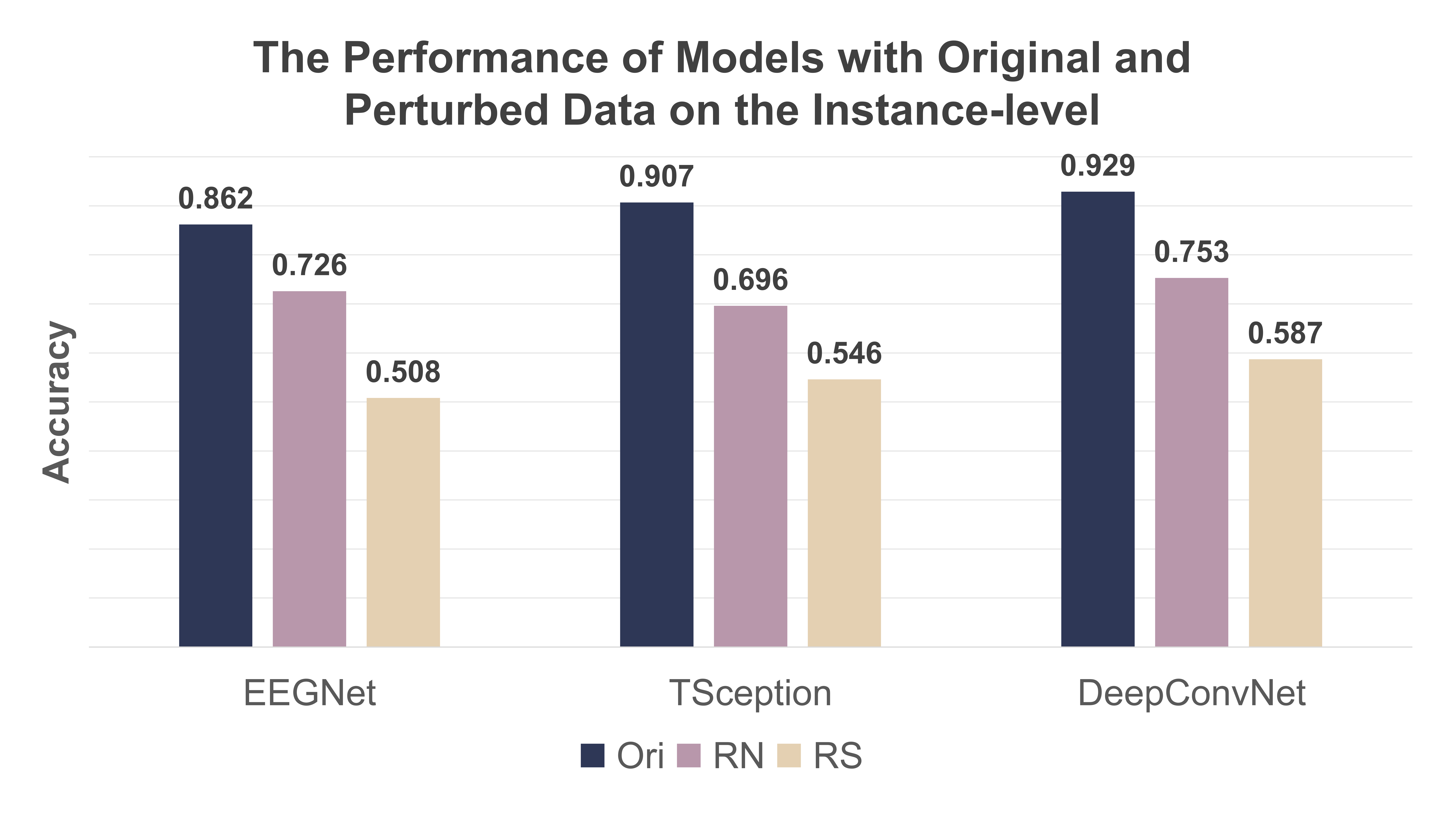}}
\caption{The result of instance-level performance. The accuracies of models using different input data are listed in the figure. Ori: original data; RN: removing non-salient components; RS: removing salient components.}
\label{fig8}
\end{figure}

To validate the instance-level capture of our method, we experiment with a similar procedure to the previous experiment. However, the averaging operation is omitted. Moreover, the dividing operation and the removal and feed-in game are performed on each instance rather than on the group. The result is shown in Fig.~\ref{fig8}. The performance of three models suffers a more significant impact when the salient components are removed. It demonstrates that our method can effectively capture the instance-level saliency.

\subsection{Saliency Validation on Unseen Subjects for Subject-independent Models}
\begin{figure*}[ht]
\centerline{\includegraphics[width=\textwidth]{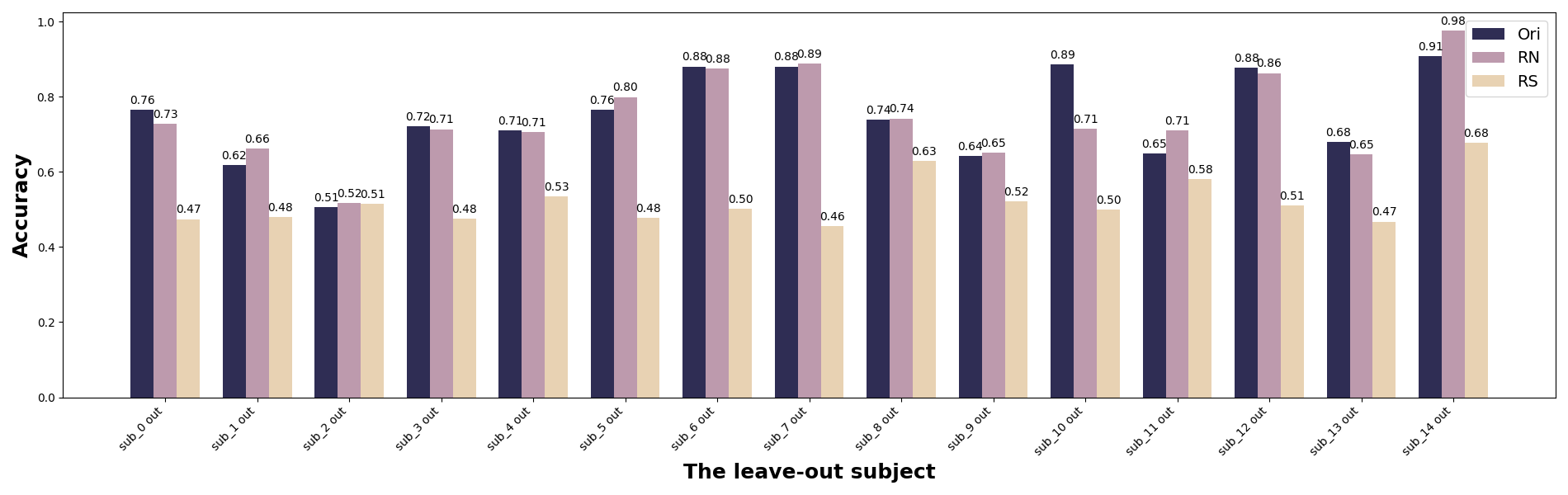}}
\caption{The performance of subject-independent models on unseen (leave-out) subject using original and perturbed data. Ori: original data; RN: removing non-salient components; RS: removing salient components.}
\label{fig9}
\end{figure*}
In practical applications, EEG models trained on the training set need to predict data from unseen subjects. Typically, due to the issue of domain shift, model performance on unseen data is often unstable. Therefore, it is necessary to explain how the model understand data from unseen subjects. This can help researchers identify challenges in building subject-independent models or localize neuro-cognitive features of different subjects. 

To this end, we adopt a leave-one-subject-out training paradigm to train the model, where each subject is sequentially excluded from the training data as unseen data. And then, we use the proposed method to capture the saliency of the samples from the unseen subject (leave-out subject in training process) on each model.

For validation, we adopt a strategy similar to previous experiments. For each sample from unseen subjects, we generate a saliency map. Then, all samples from unseen subjects are averaged to produce a group-level saliency map. Then, we use the feed-in and removal game to validate the effectiveness of the captured saliency, the results are shown in Fig.~\ref{fig9}.

As observed, except for the $sub\_2\ out$, all the listed models experience a significant drop in accuracy when using data with salient components removed, whereas the accuracy decline is less pronounced when non-salient components are removed. This result demonstrates that our method can effectively capture saliency when the data is unseen by the model. As for the $sub\_2\ out$, the model's accuracy using original data, 0.51, is very limited for a binary classification model. The incapability of the model might hinder the capture of saliency in the data.

Notably, in some cases, we observed improved model performance when non-salient components are removed from unseen data compared to using the original data. This indicates enhanced generalization to unseen subjects. We propose two possible explanations for this observation. First, features with low scores in the saliency map may be irrelevant or noisy, contributing little to the model's predictions. Removing these noisy features purifies the input data, thereby improving prediction accuracy. Second, the subject-independent model could overfit on some specific features. Thus, the prediction on unseen data is not sensitive to these features. When non-significant features are removed, these features that cause overfitting are also removed, resulting in a rise in the model's ability to generalize.

\begin{figure}[htbp]
    \centering
    \begin{subfigure}[b]{0.5\linewidth}
        \centering
        \includegraphics[width=\linewidth]{fig5_1.png}
    \end{subfigure}
    \hfill
    \begin{subfigure}[b]{0.5\linewidth}
        \centering
        \includegraphics[width=\linewidth]{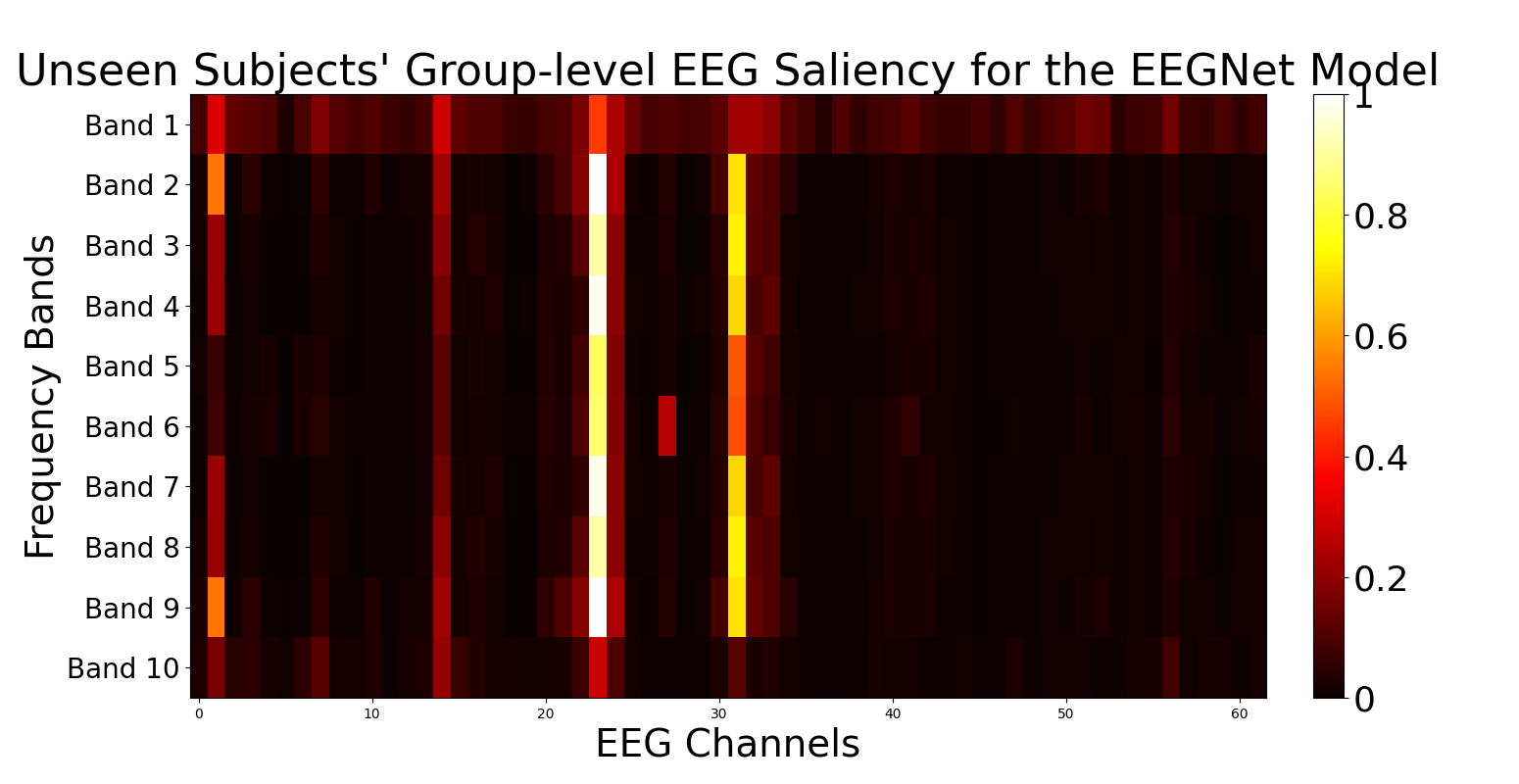}
    \end{subfigure}
    \caption{The comparison of unseen subjects' group-level saliency map and the group-level saliency map for seen subjects.}
    \label{fig10}
\end{figure}

To intuitively illustrate the generalization behavior of subject-independent models on unseen subjects, we visualize the group-level saliency map for these subjects by averaging the saliency maps of all samples from these subjects and their corresponding subject-independent models. Specifically, let $m_i \in [0,1]^{N_i \times C \times F}, i \in S$ denote the saliency maps for unseen subject $i$ on the subject-independent model $Sub\_i\ out$, where $S$ represents the subjects and $N_i$ denotes the number of samples for subject $i$. All $m_i$ are concatenated and averaged across the batch dimension to generate the group-level saliency map. This group-level saliency map for unseen subjects is compared with the group-level saliency map generated from the model trained on all subjects. The result, shown in Fig.~\ref{fig10}, reveals that the frequency components highlighted by the group-level saliency map for unseen subjects differ from those in the map obtained from seen subjects. This variation might indicate differences in the model's behavior when generalizing to unseen subjects.

In current EEG deep learning research, the generalization of subject-independent models to unseen subjects remains a challenging issue. This observation highlights the potential of explaining deep learning models in the frequency domain to reveal neuro-cognitive mechanisms and advance the application of EEG-based deep learning.

\subsection{The Effect of Target Alignment Loss Analysis}
The target alignment loss in the optimization objective explicitly encourages the perturbation to simulate the distribution of original samples, controlled by the hyper-parameter $\lambda$. It is supposed to reduce the discrepancy between the distributions of the original samples and the perturbed samples. Here, we implement an experiment using EEGNet model to analyze its effect and sensitivity to the hyper-parameter $\lambda$. To investigate its influence, we adopt kernel density estimation (KDE) to compare the two distributions. The log-likelihood is calculated to quantify the discrepancy. Moreover, we also calculate the drop in the model's performance to see its influence on group-level saliency validation. The result can be seen in Table~\ref{tab2}.

\begin{table}[h] 
\centering 
\caption{Target Alignment Loss Sensitivity Analysis} 
\label{tab2} 
\begin{tabular}{|c|c|c|}
\hline
\textbf{$\lambda$} & \textbf{KDE score} & \textbf{Performance Drop} \\
\hline
0.5 & -77.63 & 0.351 \\
\hline
0.05 & -77.23 & 0.364 \\
\hline
0.005 & -80.87 & 0.367 \\
\hline
0.0005 & -82.66 & 0.361 \\
\hline
0 (removed) & -90.66 & 0.358 \\
\hline
\end{tabular}
\end{table}

As we can see, the KDE score increases when the $\lambda$ increases from 0 to 0.05. It indicates that the likelihood of perturbed samples originating from the original distribution increases. This fact substantiates the effect of the target alignment loss on reducing distribution discrepancy. However, when the $\lambda$ increases from 0.05 to 0.5, we observe that the KDE score shows a slight decrease. Reducing the distance between perturbations and all samples in the target clusters causes the perturbations to approach the specific centroids. An excessively large $\lambda$ may cause the generated perturbation to narrow its distribution range and overly concentrate around the centroids, thereby failing to accurately simulate the overall distribution of the original samples. Additionally, although the model's performance drop is not sensitive to $\lambda$, overly emphasizing the target alignment loss can still slightly affect the effectiveness of the saliency. Therefore, we choose 0.05 as the value for $\lambda$.

\subsection{Ablation Study}
We further explore the roles of crucial components in our method. The effect of regularized terms $\Vert M \Vert_1 + \Vert {x}^{f,r} \Vert_1$ is investigated by removing it from the objective. Moreover, we validate the effect of two-branch architeture of perturbation generator network by replacing it with a one-branch architecture. The real and imaginary parts are concatenated to form the input, and the perturbation is learned by a single linear layer. Besides, the use of linear layer is also tested by replacing it with a one-layer GRU used in~\citep{r46,r48}. This experiment is implemented using the EEGNet model. The results are shown in Table~\ref{tab3}.
\begin{table}[h] 
\centering 
\caption{Ablation Study Result} 
\label{tab3} 
\begin{tabularx}{\linewidth}{|X|X|X|X|X|}
\hline
\textbf{Ablation Methods} & \textbf{Group-level RN} & \textbf{Group-level RS} & \textbf{Instance-level RN} & \textbf{Instance-level RS} \\
\hline
Our Method & 0.849 & 0.485 & 0.726 & 0.508\\
\hline
No regularized terms & 0.530 & 0.731 & 0.59 & 0.508\\
\hline
One-branch perturbation & 0.833 & 0.550 & 0.706 & 0.522\\
\hline
GRU perturbation & 0.823 & 0.495 & 0.706 & 0.588 \\
\hline
\end{tabularx}
\end{table}

All ablation experiment methods have suffered various degrees of validation loss. This indicates the effectiveness of the various modules included in our method. Moreover, during the experiments, we observed that using GRU not only fails to facilitate saliency capture but also introduces additional computational complexity compared to the structure of dense layers. Moreover, assuming the size of our input spectral samples is $D \times T$, the computational complexity of the generator we propose is $O(T\times T)$, whereas the GRU-based generator has a computational complexity of $O(6 \times D^2 \times T)$. We calculated the FLOPs for both types of generators, with our proposed method yielding 320K, and the GRU-based generator reaching 18.4512M. Therefore, our method not only demonstrates superiority in effectiveness but also significantly reduces computational costs. 

\section{Conclusion}
In this study, we introduce a novel mask perturbation method to expose the saliency for EEG-based end-to-end deep learning models. Our approach demonstrates significant advantages over existing methods. To address the out-of-distribution problem, we propose a target alignment loss that simulates the original distribution of EEG data. Additionally, we present a simple yet effective perturbation generator that defines perturbations in the frequency domain for EEG data.

Currently, capturing instance-level saliency is less effective compared to group-level saliency. Future work can focus on addressing this limitation. Experimental results indicate that our proposed method still faces challenges in aligning perturbations with the original samples. The effectiveness of the proposed method is dependent on the careful selection of hyper-parameters. Future studies should explore automatic methods for distribution alignment as a potential solution.

\section*{Acknowledgments}
This research is partly funded by the China Mobile Research Fund of Chinese Ministry of Education (Grant No. KEH2310029). This work is also supported by the Joint Lab on Networked AI Edge Computing Fudan University-Changan.

\section*{Declaration of Competing Interest}
The authors declare that they have no known competing financial interests or personal relationships that could have appeared to influence the work reported in this paper.

\bibliographystyle{elsarticle-num-names} 
\bibliography{ref}

\begin{thebibliography}{46}
\expandafter\ifx\csname natexlab\endcsname\relax\def\natexlab#1{#1}\fi
\providecommand{\url}[1]{\texttt{#1}}
\providecommand{\href}[2]{#2}
\providecommand{\path}[1]{#1}
\providecommand{\DOIprefix}{doi:}
\providecommand{\ArXivprefix}{arXiv:}
\providecommand{\URLprefix}{URL: }
\providecommand{\Pubmedprefix}{pmid:}
\providecommand{\doi}[1]{\href{http://dx.doi.org/#1}{\path{#1}}}
\providecommand{\Pubmed}[1]{\href{pmid:#1}{\path{#1}}}
\providecommand{\bibinfo}[2]{#2}
\ifx\xfnm\relax \def\xfnm[#1]{\unskip,\space#1}\fi
\bibitem[{Roy et~al.(2019)Roy, Banville, Albuquerque, Gramfort, Falk, and Faubert}]{r1}
\bibinfo{author}{Y.~Roy}, \bibinfo{author}{H.~Banville}, \bibinfo{author}{I.~Albuquerque}, \bibinfo{author}{A.~Gramfort}, \bibinfo{author}{T.~H. Falk}, \bibinfo{author}{J.~Faubert},
\newblock \bibinfo{title}{Deep learning-based electroencephalography analysis: a systematic review},
\newblock \bibinfo{journal}{Journal of neural engineering} \bibinfo{volume}{16} (\bibinfo{year}{2019}) \bibinfo{pages}{051001}.
\bibitem[{Jafari et~al.(2023)Jafari, Shoeibi, Khodatars, Bagherzadeh, Shalbaf, Garc{\'\i}a, Gorriz, and Acharya}]{r15}
\bibinfo{author}{M.~Jafari}, \bibinfo{author}{A.~Shoeibi}, \bibinfo{author}{M.~Khodatars}, \bibinfo{author}{S.~Bagherzadeh}, \bibinfo{author}{A.~Shalbaf}, \bibinfo{author}{D.~L. Garc{\'\i}a}, \bibinfo{author}{J.~M. Gorriz}, \bibinfo{author}{U.~R. Acharya},
\newblock \bibinfo{title}{Emotion recognition in eeg signals using deep learning methods: A review},
\newblock \bibinfo{journal}{Computers in Biology and Medicine}  (\bibinfo{year}{2023}) \bibinfo{pages}{107450}.
\bibitem[{Jas et~al.(2017)Jas, Engemann, Bekhti, Raimondo, and Gramfort}]{r2}
\bibinfo{author}{M.~Jas}, \bibinfo{author}{D.~A. Engemann}, \bibinfo{author}{Y.~Bekhti}, \bibinfo{author}{F.~Raimondo}, \bibinfo{author}{A.~Gramfort},
\newblock \bibinfo{title}{Autoreject: Automated artifact rejection for meg and eeg data},
\newblock \bibinfo{journal}{NeuroImage} \bibinfo{volume}{159} (\bibinfo{year}{2017}) \bibinfo{pages}{417--429}.
\bibitem[{Craik et~al.(2019)Craik, He, and Contreras-Vidal}]{r3}
\bibinfo{author}{A.~Craik}, \bibinfo{author}{Y.~He}, \bibinfo{author}{J.~L. Contreras-Vidal},
\newblock \bibinfo{title}{Deep learning for electroencephalogram (eeg) classification tasks: a review},
\newblock \bibinfo{journal}{Journal of neural engineering} \bibinfo{volume}{16} (\bibinfo{year}{2019}) \bibinfo{pages}{031001}.
\bibitem[{Autthasan et~al.(2021)Autthasan, Chaisaen, Sudhawiyangkul, Kiatthaveephong, Rangpong, Dilokthanakul, Bhakdisongkhram, Phan, Guan, and Wilaiprasitporn}]{r4}
\bibinfo{author}{P.~Autthasan}, \bibinfo{author}{R.~Chaisaen}, \bibinfo{author}{T.~Sudhawiyangkul}, \bibinfo{author}{S.~Kiatthaveephong}, \bibinfo{author}{P.~Rangpong}, \bibinfo{author}{N.~Dilokthanakul}, \bibinfo{author}{G.~Bhakdisongkhram}, \bibinfo{author}{H.~Phan}, \bibinfo{author}{C.~Guan}, \bibinfo{author}{T.~Wilaiprasitporn},
\newblock \bibinfo{title}{Min2net: End-to-end multi-task learning for subject-independent motor imagery eeg classification},
\newblock \bibinfo{journal}{IEEE Transactions on Biomedical Engineering}  (\bibinfo{year}{2021}).
\bibitem[{Perslev et~al.(2019)Perslev, Jensen, Darkner, Jennum, and Igel}]{r5}
\bibinfo{author}{M.~Perslev}, \bibinfo{author}{M.~Jensen}, \bibinfo{author}{S.~Darkner}, \bibinfo{author}{P.~J. Jennum}, \bibinfo{author}{C.~Igel},
\newblock \bibinfo{title}{U-time: A fully convolutional network for time series segmentation applied to sleep staging},
\newblock \bibinfo{journal}{Advances in Neural Information Processing Systems} \bibinfo{volume}{32} (\bibinfo{year}{2019}).
\bibitem[{Su et~al.(2022)Su, Cai, Xie, Li, and Schultz}]{r6}
\bibinfo{author}{E.~Su}, \bibinfo{author}{S.~Cai}, \bibinfo{author}{L.~Xie}, \bibinfo{author}{H.~Li}, \bibinfo{author}{T.~Schultz},
\newblock \bibinfo{title}{Stanet: A spatiotemporal attention network for decoding auditory spatial attention from eeg},
\newblock \bibinfo{journal}{IEEE Transactions on Biomedical Engineering}  (\bibinfo{year}{2022}).
\bibitem[{Eldele et~al.(2021)Eldele, Chen, Liu, Wu, Kwoh, Li, and Guan}]{r7}
\bibinfo{author}{E.~Eldele}, \bibinfo{author}{Z.~Chen}, \bibinfo{author}{C.~Liu}, \bibinfo{author}{M.~Wu}, \bibinfo{author}{C.-K. Kwoh}, \bibinfo{author}{X.~Li}, \bibinfo{author}{C.~Guan},
\newblock \bibinfo{title}{An attention-based deep learning approach for sleep stage classification with single-channel eeg},
\newblock \bibinfo{journal}{IEEE Transactions on Neural Systems and Rehabilitation Engineering} \bibinfo{volume}{29} (\bibinfo{year}{2021}) \bibinfo{pages}{809--818}.
\bibitem[{Lawhern et~al.(2018)Lawhern, Solon, Waytowich, Gordon, Hung, and Lance}]{r8}
\bibinfo{author}{V.~J. Lawhern}, \bibinfo{author}{A.~J. Solon}, \bibinfo{author}{N.~R. Waytowich}, \bibinfo{author}{S.~M. Gordon}, \bibinfo{author}{C.~P. Hung}, \bibinfo{author}{B.~J. Lance},
\newblock \bibinfo{title}{Eegnet: a compact convolutional neural network for eeg-based brain--computer interfaces},
\newblock \bibinfo{journal}{Journal of neural engineering} \bibinfo{volume}{15} (\bibinfo{year}{2018}) \bibinfo{pages}{056013}.
\bibitem[{Liang et~al.(2021)Liang, Zhou, Zhang, Li, Huang, Zhang, and Ishii}]{r9}
\bibinfo{author}{Z.~Liang}, \bibinfo{author}{R.~Zhou}, \bibinfo{author}{L.~Zhang}, \bibinfo{author}{L.~Li}, \bibinfo{author}{G.~Huang}, \bibinfo{author}{Z.~Zhang}, \bibinfo{author}{S.~Ishii},
\newblock \bibinfo{title}{Eegfusenet: Hybrid unsupervised deep feature characterization and fusion for high-dimensional eeg with an application to emotion recognition},
\newblock \bibinfo{journal}{IEEE Transactions on Neural Systems and Rehabilitation Engineering} \bibinfo{volume}{29} (\bibinfo{year}{2021}) \bibinfo{pages}{1913--1925}.
\bibitem[{Zheng and Lu(2015)}]{r10}
\bibinfo{author}{W.-L. Zheng}, \bibinfo{author}{B.-L. Lu},
\newblock \bibinfo{title}{Investigating critical frequency bands and channels for eeg-based emotion recognition with deep neural networks},
\newblock \bibinfo{journal}{IEEE Transactions on autonomous mental development} \bibinfo{volume}{7} (\bibinfo{year}{2015}) \bibinfo{pages}{162--175}.
\bibitem[{Li et~al.(2021)Li, Wang, and Lu}]{r11}
\bibinfo{author}{R.~Li}, \bibinfo{author}{Y.~Wang}, \bibinfo{author}{B.-L. Lu},
\newblock \bibinfo{title}{A multi-domain adaptive graph convolutional network for eeg-based emotion recognition},
\newblock in: \bibinfo{booktitle}{Proceedings of the 29th ACM International Conference on Multimedia}, \bibinfo{year}{2021}, pp. \bibinfo{pages}{5565--5573}.
\bibitem[{Duan et~al.(2013)Duan, Zhu, and Lu}]{r12}
\bibinfo{author}{R.-N. Duan}, \bibinfo{author}{J.-Y. Zhu}, \bibinfo{author}{B.-L. Lu},
\newblock \bibinfo{title}{Differential entropy feature for eeg-based emotion classification},
\newblock in: \bibinfo{booktitle}{2013 6th International IEEE/EMBS Conference on Neural Engineering (NER)}, \bibinfo{organization}{IEEE}, \bibinfo{year}{2013}, pp. \bibinfo{pages}{81--84}.
\bibitem[{r13(2022)}]{r13}
\bibinfo{title}{Time-frequency analysis methods and their application in developmental eeg data},
\newblock \bibinfo{journal}{Developmental Cognitive Neuroscience} \bibinfo{volume}{54} (\bibinfo{year}{2022}) \bibinfo{pages}{101067}.
\bibitem[{Wang et~al.(2022)Wang, Zhu, Chen, Li, and Song}]{r14}
\bibinfo{author}{H.~Wang}, \bibinfo{author}{X.~Zhu}, \bibinfo{author}{T.~Chen}, \bibinfo{author}{C.~Li}, \bibinfo{author}{L.~Song},
\newblock \bibinfo{title}{Rethinking saliency map: A context-aware perturbation method to explain eeg-based deep learning model},
\newblock \bibinfo{journal}{IEEE Transactions on Biomedical Engineering}  (\bibinfo{year}{2022}).
\bibitem[{Fellous et~al.(2019)Fellous, Sapiro, Rossi, Mayberg, and Ferrante}]{r16}
\bibinfo{author}{J.-M. Fellous}, \bibinfo{author}{G.~Sapiro}, \bibinfo{author}{A.~Rossi}, \bibinfo{author}{H.~Mayberg}, \bibinfo{author}{M.~Ferrante},
\newblock \bibinfo{title}{Explainable artificial intelligence for neuroscience: behavioral neurostimulation},
\newblock \bibinfo{journal}{Frontiers in neuroscience} \bibinfo{volume}{13} (\bibinfo{year}{2019}) \bibinfo{pages}{490966}.
\bibitem[{Nahmias and Kontson(2020)}]{r23}
\bibinfo{author}{D.~O. Nahmias}, \bibinfo{author}{K.~L. Kontson},
\newblock \bibinfo{title}{Easy perturbation eeg algorithm for spectral importance (easypeasi) a simple method to identify important spectral features of eeg in deep learning models},
\newblock in: \bibinfo{booktitle}{Proceedings of the 26th ACM SIGKDD International Conference on Knowledge Discovery \& Data Mining}, \bibinfo{year}{2020}, pp. \bibinfo{pages}{2398--2406}.
\bibitem[{Farahat et~al.(2019)Farahat, Reichert, Sweeney-Reed, and Hinrichs}]{r17}
\bibinfo{author}{A.~Farahat}, \bibinfo{author}{C.~Reichert}, \bibinfo{author}{C.~M. Sweeney-Reed}, \bibinfo{author}{H.~Hinrichs},
\newblock \bibinfo{title}{Convolutional neural networks for decoding of covert attention focus and saliency maps for eeg feature visualization},
\newblock \bibinfo{journal}{Journal of neural engineering} \bibinfo{volume}{16} (\bibinfo{year}{2019}) \bibinfo{pages}{066010}.
\bibitem[{Zang et~al.(2021)Zang, Lin, Liu, and Gao}]{r18}
\bibinfo{author}{B.~Zang}, \bibinfo{author}{Y.~Lin}, \bibinfo{author}{Z.~Liu}, \bibinfo{author}{X.~Gao},
\newblock \bibinfo{title}{A deep learning method for single-trial eeg classification in rsvp task based on spatiotemporal features of erps},
\newblock \bibinfo{journal}{Journal of Neural Engineering} \bibinfo{volume}{18} (\bibinfo{year}{2021}) \bibinfo{pages}{0460c8}.
\bibitem[{Lin et~al.(2021)Lin, Jia, Li, Li, Qian, Li, Pan, and Ji}]{r19}
\bibinfo{author}{P.-J. Lin}, \bibinfo{author}{T.~Jia}, \bibinfo{author}{C.~Li}, \bibinfo{author}{T.~Li}, \bibinfo{author}{C.~Qian}, \bibinfo{author}{Z.~Li}, \bibinfo{author}{Y.~Pan}, \bibinfo{author}{L.~Ji},
\newblock \bibinfo{title}{Cnn-based prognosis of bci rehabilitation using eeg from first session bci training},
\newblock \bibinfo{journal}{IEEE Transactions on Neural Systems and Rehabilitation Engineering} \bibinfo{volume}{29} (\bibinfo{year}{2021}) \bibinfo{pages}{1936--1943}.
\bibitem[{Ding et~al.(2021)Ding, Robinson, Zhang, Zeng, and Guan}]{r20}
\bibinfo{author}{Y.~Ding}, \bibinfo{author}{N.~Robinson}, \bibinfo{author}{S.~Zhang}, \bibinfo{author}{Q.~Zeng}, \bibinfo{author}{C.~Guan},
\newblock \bibinfo{title}{Tsception: Capturing temporal dynamics and spatial asymmetry from eeg for emotion recognition},
\newblock \bibinfo{journal}{arXiv preprint arXiv:2104.02935}  (\bibinfo{year}{2021}).
\bibitem[{Schirrmeister et~al.(2017)Schirrmeister, Springenberg, Fiederer, Glasstetter, Eggensperger, Tangermann, Hutter, Burgard, and Ball}]{r21}
\bibinfo{author}{R.~T. Schirrmeister}, \bibinfo{author}{J.~T. Springenberg}, \bibinfo{author}{L.~D.~J. Fiederer}, \bibinfo{author}{M.~Glasstetter}, \bibinfo{author}{K.~Eggensperger}, \bibinfo{author}{M.~Tangermann}, \bibinfo{author}{F.~Hutter}, \bibinfo{author}{W.~Burgard}, \bibinfo{author}{T.~Ball},
\newblock \bibinfo{title}{Deep learning with convolutional neural networks for eeg decoding and visualization},
\newblock \bibinfo{journal}{Human brain mapping} \bibinfo{volume}{38} (\bibinfo{year}{2017}) \bibinfo{pages}{5391--5420}.
\bibitem[{Hartmann et~al.(2018)Hartmann, Schirrmeister, and Ball}]{r22}
\bibinfo{author}{K.~G. Hartmann}, \bibinfo{author}{R.~T. Schirrmeister}, \bibinfo{author}{T.~Ball},
\newblock \bibinfo{title}{Hierarchical internal representation of spectral features in deep convolutional networks trained for eeg decoding},
\newblock in: \bibinfo{booktitle}{2018 6th International Conference on Brain-Computer Interface (BCI)}, \bibinfo{organization}{IEEE}, \bibinfo{year}{2018}, pp. \bibinfo{pages}{1--6}.
\bibitem[{Miskovic et~al.(2011)Miskovic, Moscovitch, Santesso, McCabe, Antony, and Schmidt}]{r28}
\bibinfo{author}{V.~Miskovic}, \bibinfo{author}{D.~A. Moscovitch}, \bibinfo{author}{D.~L. Santesso}, \bibinfo{author}{R.~E. McCabe}, \bibinfo{author}{M.~M. Antony}, \bibinfo{author}{L.~A. Schmidt},
\newblock \bibinfo{title}{Changes in eeg cross-frequency coupling during cognitive behavioral therapy for social anxiety disorder},
\newblock \bibinfo{journal}{Psychological science} \bibinfo{volume}{22} (\bibinfo{year}{2011}) \bibinfo{pages}{507--516}.
\bibitem[{Whittingstall and Logothetis(2009)}]{r29}
\bibinfo{author}{K.~Whittingstall}, \bibinfo{author}{N.~K. Logothetis},
\newblock \bibinfo{title}{Frequency-band coupling in surface eeg reflects spiking activity in monkey visual cortex},
\newblock \bibinfo{journal}{Neuron} \bibinfo{volume}{64} (\bibinfo{year}{2009}) \bibinfo{pages}{281--289}.
\bibitem[{Canolty and Knight(2010)}]{r30}
\bibinfo{author}{R.~T. Canolty}, \bibinfo{author}{R.~T. Knight},
\newblock \bibinfo{title}{The functional role of cross-frequency coupling},
\newblock \bibinfo{journal}{Trends in cognitive sciences} \bibinfo{volume}{14} (\bibinfo{year}{2010}) \bibinfo{pages}{506--515}.
\bibitem[{Fong and Vedaldi(2017)}]{r24}
\bibinfo{author}{R.~C. Fong}, \bibinfo{author}{A.~Vedaldi},
\newblock \bibinfo{title}{Interpretable explanations of black boxes by meaningful perturbation},
\newblock in: \bibinfo{booktitle}{Proceedings of the IEEE international conference on computer vision}, \bibinfo{year}{2017}, pp. \bibinfo{pages}{3429--3437}.
\bibitem[{Dabkowski and Gal(2017)}]{r25}
\bibinfo{author}{P.~Dabkowski}, \bibinfo{author}{Y.~Gal},
\newblock \bibinfo{title}{Real time image saliency for black box classifiers},
\newblock \bibinfo{journal}{Advances in neural information processing systems} \bibinfo{volume}{30} (\bibinfo{year}{2017}).
\bibitem[{Fong et~al.(2019)Fong, Patrick, and Vedaldi}]{r26}
\bibinfo{author}{R.~Fong}, \bibinfo{author}{M.~Patrick}, \bibinfo{author}{A.~Vedaldi},
\newblock \bibinfo{title}{Understanding deep networks via extremal perturbations and smooth masks},
\newblock in: \bibinfo{booktitle}{Proceedings of the IEEE/CVF international conference on computer vision}, \bibinfo{year}{2019}, pp. \bibinfo{pages}{2950--2958}.
\bibitem[{Crabb{\'e} and Van Der~Schaar(2021)}]{r27}
\bibinfo{author}{J.~Crabb{\'e}}, \bibinfo{author}{M.~Van Der~Schaar},
\newblock \bibinfo{title}{Explaining time series predictions with dynamic masks},
\newblock in: \bibinfo{booktitle}{International Conference on Machine Learning}, \bibinfo{organization}{PMLR}, \bibinfo{year}{2021}, pp. \bibinfo{pages}{2166--2177}.
\bibitem[{Nguyen et~al.(2015)Nguyen, Yosinski, and Clune}]{r32}
\bibinfo{author}{A.~Nguyen}, \bibinfo{author}{J.~Yosinski}, \bibinfo{author}{J.~Clune},
\newblock \bibinfo{title}{Deep neural networks are easily fooled: High confidence predictions for unrecognizable images},
\newblock in: \bibinfo{booktitle}{Proceedings of the IEEE conference on computer vision and pattern recognition}, \bibinfo{year}{2015}, pp. \bibinfo{pages}{427--436}.
\bibitem[{Mahendran and Vedaldi(2015)}]{r31}
\bibinfo{author}{A.~Mahendran}, \bibinfo{author}{A.~Vedaldi},
\newblock \bibinfo{title}{Understanding deep image representations by inverting them},
\newblock in: \bibinfo{booktitle}{Proceedings of the IEEE conference on computer vision and pattern recognition}, \bibinfo{year}{2015}, pp. \bibinfo{pages}{5188--5196}.
\bibitem[{Al-Nafjan et~al.(2017)Al-Nafjan, Hosny, Al-Ohali, and Al-Wabil}]{r33}
\bibinfo{author}{A.~Al-Nafjan}, \bibinfo{author}{M.~Hosny}, \bibinfo{author}{Y.~Al-Ohali}, \bibinfo{author}{A.~Al-Wabil},
\newblock \bibinfo{title}{Review and classification of emotion recognition based on eeg brain-computer interface system research: a systematic review},
\newblock \bibinfo{journal}{Applied Sciences} \bibinfo{volume}{7} (\bibinfo{year}{2017}) \bibinfo{pages}{1239}.
\bibitem[{Selvaraju et~al.(2017)Selvaraju, Cogswell, Das, Vedantam, Parikh, and Batra}]{r35}
\bibinfo{author}{R.~R. Selvaraju}, \bibinfo{author}{M.~Cogswell}, \bibinfo{author}{A.~Das}, \bibinfo{author}{R.~Vedantam}, \bibinfo{author}{D.~Parikh}, \bibinfo{author}{D.~Batra},
\newblock \bibinfo{title}{Grad-cam: Visual explanations from deep networks via gradient-based localization},
\newblock in: \bibinfo{booktitle}{Proceedings of the IEEE international conference on computer vision}, \bibinfo{year}{2017}, pp. \bibinfo{pages}{618--626}.
\bibitem[{Li et~al.(2020)Li, Yang, Li, Chen, and Du}]{r36}
\bibinfo{author}{Y.~Li}, \bibinfo{author}{H.~Yang}, \bibinfo{author}{J.~Li}, \bibinfo{author}{D.~Chen}, \bibinfo{author}{M.~Du},
\newblock \bibinfo{title}{Eeg-based intention recognition with deep recurrent-convolution neural network: Performance and channel selection by grad-cam},
\newblock \bibinfo{journal}{Neurocomputing} \bibinfo{volume}{415} (\bibinfo{year}{2020}) \bibinfo{pages}{225--233}.
\bibitem[{Jonas et~al.(2019)Jonas, Rossetti, Oddo, Jenni, Favaro, and Zubler}]{r37}
\bibinfo{author}{S.~Jonas}, \bibinfo{author}{A.~O. Rossetti}, \bibinfo{author}{M.~Oddo}, \bibinfo{author}{S.~Jenni}, \bibinfo{author}{P.~Favaro}, \bibinfo{author}{F.~Zubler},
\newblock \bibinfo{title}{Eeg-based outcome prediction after cardiac arrest with convolutional neural networks: Performance and visualization of discriminative features},
\newblock \bibinfo{journal}{Human brain mapping} \bibinfo{volume}{40} (\bibinfo{year}{2019}) \bibinfo{pages}{4606--4617}.
\bibitem[{Yan et~al.(2021)Yan, Zhou, Huang, Cheng, and Kuang}]{r38}
\bibinfo{author}{Y.~Yan}, \bibinfo{author}{H.~Zhou}, \bibinfo{author}{L.~Huang}, \bibinfo{author}{X.~Cheng}, \bibinfo{author}{S.~Kuang},
\newblock \bibinfo{title}{A novel two-stage refine filtering method for eeg-based motor imagery classification},
\newblock \bibinfo{journal}{Frontiers in neuroscience} \bibinfo{volume}{15} (\bibinfo{year}{2021}).
\bibitem[{Zeiler and Fergus(2014)}]{r39}
\bibinfo{author}{M.~D. Zeiler}, \bibinfo{author}{R.~Fergus},
\newblock \bibinfo{title}{Visualizing and understanding convolutional networks},
\newblock in: \bibinfo{booktitle}{European conference on computer vision}, \bibinfo{organization}{Springer}, \bibinfo{year}{2014}, pp. \bibinfo{pages}{818--833}.
\bibitem[{Phillips et~al.(2019)Phillips, Goh, and Hodas}]{r40}
\bibinfo{author}{L.~Phillips}, \bibinfo{author}{G.~Goh}, \bibinfo{author}{N.~Hodas},
\newblock \bibinfo{title}{Explanatory masks for neural network interpretability},
\newblock \bibinfo{journal}{arXiv preprint arXiv:1911.06876}  (\bibinfo{year}{2019}).
\bibitem[{Ho et~al.(2021)Ho, Aczon, Ledbetter, and Wetzel}]{r41}
\bibinfo{author}{L.~V. Ho}, \bibinfo{author}{M.~Aczon}, \bibinfo{author}{D.~Ledbetter}, \bibinfo{author}{R.~Wetzel},
\newblock \bibinfo{title}{Interpreting a recurrent neural network’s predictions of icu mortality risk},
\newblock \bibinfo{journal}{Journal of Biomedical Informatics} \bibinfo{volume}{114} (\bibinfo{year}{2021}) \bibinfo{pages}{103672}.
\bibitem[{Enguehard(2023)}]{r46}
\bibinfo{author}{J.~Enguehard},
\newblock \bibinfo{title}{Learning perturbations to explain time series predictions},
\newblock in: \bibinfo{booktitle}{International Conference on Machine Learning}, \bibinfo{organization}{PMLR}, \bibinfo{year}{2023}, pp. \bibinfo{pages}{9329--9342}.
\bibitem[{Liu et~al.(2024)Liu, Zhang, Wang, Wang, Luo, Du, Wu, Wang, Chen, Fan et~al.}]{r48}
\bibinfo{author}{Z.~Liu}, \bibinfo{author}{Y.~Zhang}, \bibinfo{author}{T.~Wang}, \bibinfo{author}{Z.~Wang}, \bibinfo{author}{D.~Luo}, \bibinfo{author}{M.~Du}, \bibinfo{author}{M.~Wu}, \bibinfo{author}{Y.~Wang}, \bibinfo{author}{C.~Chen}, \bibinfo{author}{L.~Fan}, et~al.,
\newblock \bibinfo{title}{Explaining time series via contrastive and locally sparse perturbations},
\newblock \bibinfo{journal}{arXiv preprint arXiv:2401.08552}  (\bibinfo{year}{2024}).
\bibitem[{Joo et~al.(2023)Joo, Quan, Trang, Kim, and Woo}]{r42}
\bibinfo{author}{H.~Joo}, \bibinfo{author}{L.~D.~A. Quan}, \bibinfo{author}{L.~T. Trang}, \bibinfo{author}{D.~Kim}, \bibinfo{author}{J.~Woo},
\newblock \bibinfo{title}{Group-level interpretation of electroencephalography signals using compact convolutional neural networks},
\newblock \bibinfo{journal}{IEEE Access} \bibinfo{volume}{11} (\bibinfo{year}{2023}) \bibinfo{pages}{114992--115001}.
\bibitem[{Bigdely-Shamlo et~al.(2013)Bigdely-Shamlo, Mullen, Kreutz-Delgado, and Makeig}]{r43}
\bibinfo{author}{N.~Bigdely-Shamlo}, \bibinfo{author}{T.~Mullen}, \bibinfo{author}{K.~Kreutz-Delgado}, \bibinfo{author}{S.~Makeig},
\newblock \bibinfo{title}{Measure projection analysis: a probabilistic approach to eeg source comparison and multi-subject inference},
\newblock \bibinfo{journal}{Neuroimage} \bibinfo{volume}{72} (\bibinfo{year}{2013}) \bibinfo{pages}{287--303}.
\bibitem[{Tanaka(2020)}]{r44}
\bibinfo{author}{H.~Tanaka},
\newblock \bibinfo{title}{Group task-related component analysis (gtrca): A multivariate method for inter-trial reproducibility and inter-subject similarity maximization for eeg data analysis},
\newblock \bibinfo{journal}{Scientific reports} \bibinfo{volume}{10} (\bibinfo{year}{2020}) \bibinfo{pages}{84}.
\bibitem[{Csaky et~al.(2023)Csaky, Van~Es, Jones, and Woolrich}]{r45}
\bibinfo{author}{R.~Csaky}, \bibinfo{author}{M.~W. Van~Es}, \bibinfo{author}{O.~P. Jones}, \bibinfo{author}{M.~Woolrich},
\newblock \bibinfo{title}{Group-level brain decoding with deep learning},
\newblock \bibinfo{journal}{Human Brain Mapping} \bibinfo{volume}{44} (\bibinfo{year}{2023}) \bibinfo{pages}{6105--6119}.

\end{thebibliography}

\end{document}